\documentclass[sigconf]{acmart}

\copyrightyear{2026}
\acmYear{2026}
\setcopyright{cc}
\setcctype{by}
\acmConference[CIKM '26]{Proceedings of the 35th ACM International Conference on Information and Knowledge Management}{November 07--11, 2026}{Rome, Italy}
\acmBooktitle{Proceedings of the 35th ACM International Conference on Information and Knowledge Management (CIKM '26), November 07--11, 2026, Rome, Italy}
\acmDOI{10.1145/3799682.3840970}
\acmISBN{979-8-4007-2539-5/2026/11}

\usepackage{enumitem}
\usepackage{multirow}
\usepackage{multicol}
\usepackage{hyperref}
\usepackage{amsmath}
\usepackage{xspace}
\usepackage{mathtools}
\usepackage{array}
\usepackage[table]{xcolor}
\usepackage[labelfont=bf]{caption}
\usepackage{tcolorbox}
\usepackage{subcaption} 
\usepackage{algorithm}
\usepackage{algpseudocode}
\usepackage{amsthm}
\usepackage{lipsum}
\usepackage{rotating}

\usepackage{booktabs}
\usepackage{subcaption}
\usepackage{thmtools}
\usepackage{thm-restate}

\newtheorem{problem}{Problem}

\newcommand{\graph}{\ensuremath{\mathcal{G}}\xspace}
\newcommand{\nodes}{\ensuremath{\mathcal{V}}\xspace}
\newcommand{\edges}{\ensuremath{\mathcal{E}}\xspace}
\newcommand{\innateopinions}{\ensuremath{\mathbf{s}}\xspace}
\newcommand{\expressedopinions}{\ensuremath{\mathbf{z}}\xspace}

\newcommand{\node}{\ensuremath{u}\xspace}

\newcommand{\innateopinion}{\ensuremath{s}\xspace}
\newcommand{\expressedopinion}{\ensuremath{z}\xspace}

\newcommand{\content}{\ensuremath{c}\xspace}

\newcommand{\nnodes}{\ensuremath{n}\xspace}
\newcommand{\nedges}{\ensuremath{m}\xspace}

\newcommand{\prompt}{\ensuremath{q}\xspace}

\newcommand{\disruption}{\ensuremath{I}\xspace}

\newcommand{\genericvec}{\ensuremath{\mathbf{x}}\xspace}

\newcommand{\tildex}{\ensuremath{\tilde{x}}\xspace}

\usepackage{xargs}                      
\usepackage[pdftex,dvipsnames]{xcolor}  
\usepackage[colorinlistoftodos,prependcaption,textsize=tiny]{todonotes}
\newcommandx{\unsure}[2][1=]{\todo[linecolor=red,backgroundcolor=red!25,bordercolor=red,#1]{#2}}
\newcommandx{\change}[2][1=]{\todo[linecolor=blue,backgroundcolor=blue!25,bordercolor=blue,#1]{#2}}
\newcommandx{\info}[2][1=]{\todo[linecolor=OliveGreen,backgroundcolor=OliveGreen!25,bordercolor=OliveGreen,#1]{#2}}
\newcommandx{\improvement}[2][1=]{\todo[linecolor=Plum,backgroundcolor=Plum!25,bordercolor=Plum,#1]{#2}}

\begin{document}


\title{Disrupting Networks: Amplifying Social Dissensus via\\Large Language Models}


\author{Erica Coppolillo}
\affiliation{%
  \institution{ICAR-CNR}
  \city{Rende}
  \country{Italy}}
\email{ericacoppolillo@cnr.it}

\author{Giuseppe Manco}
\affiliation{%
  \institution{ICAR-CNR}
  \city{Rende}
  \country{Italy}
}
\email{giuseppe-manco@cnr.it}


\begin{abstract}
In this paper, we study how targeted content injection can strategically disrupt social networks. Specifically, we address a practical limitation of the basic variant of the Friedkin–Johnsen (FJ) model, which prevents increasing the polarization and disagreement of the network, making it unable to reproduce real-world scenarios.   We overcome such limitation by using an enriched variant of the FJ model, further showing how altering a node inherent opinion to maximize disruption. Building on these preliminary considerations, we design a reinforcement learning framework to fine-tune a Large Language Model (LLM) for generating disruption-oriented text. Experiments on synthetic and real-world data confirm that tuned LLMs can approach theoretical disruption limits. Our findings raise important considerations for content moderation, adversarial information campaigns, and generative model regulation. 
Code is publicly available.\footnote{  \url{https://github.com/EricaCoppolillo/DisruptingNetworks}}
\end{abstract}

\maketitle

\section{Introduction}
\label{sec:introduction}

Online social networks have become the primary infrastructure for information dissemination and public discourse~\cite{gallup, socialmedia}. Through complex social interactions, users influence one another~\cite{10.1145/1401890.1401897}, forming echo chambers and reinforcing beliefs, and in many settings contributing to polarization~\cite{2102141118, pranesh2024impactsocialmediapolarization}. 
In parallel, platforms have become a contested space for adversarial influence operations and coordinated manipulation, where strategic content injection can exacerbate disagreement and fracture communities~\cite{chen2020networkdisruptionmaximizingdisagreement, 10831299}.

To investigate this phenomenon, we center our analysis on \textit{social disruption}, a network-level metric that integrates polarization and disagreement~\cite{minimizing-polarization}, and we pose two central research questions:
(i) \textit{how can an agent strategically shape opinions to maximize disruption?} and
(ii) \textit{can modern generative AI systems operationalize such strategies at scale?}

\noindent{\textbf{Motivation.}} 
We build on the Friedkin--Johnsen (FJ) model~\cite{friedkin1990social}, a classical framework that interpolates between individuals innate opinions and neighbors influence via an averaging-type update. In the standard setting, the model converges to a unique steady state that is an affine combination of the innate opinions, hence remaining within their convex hull. As a consequence, it \textit{cannot} generate {repulsion} effects or amplify extremes~\cite{friedkin1990social, friedkin2011social, grabisch2020survey}. 
{This limitation is aligned with a broader literature taxonomy, where purely assimilative dynamics reduce opinion distances, while persistent polarization typically requires either similarity-biased interactions or explicitly repulsive/antagonistic mechanisms~\cite{grabisch2020survey, flache2017models}.}
In contrast, empirical studies of online debates report sustained or increasing disagreement and polarization, echo chamber formation, and bimodal opinion distributions~\cite{garimella2018quantifying, cinelli2021echo, morales2015measuring, bail2018exposure, conover2011political, del2016echo}.

\noindent{\textbf{Contributions.}} To bridge this gap, we adopt an enriched FJ model combining user susceptibility~\cite{10.1145/3219819.3219983} with signed influence weights~\cite{10591448}. Incorporating antagonistic ties is a well-established route to obtain polarization phenomena in opinion dynamics: antagonistic interaction models can yield polarization under structural-balance-type conditions~\cite{altafini2013consensus, proskurnikov2016opinion}, and recent work has explicitly studied the FJ model on signed graphs and related computational problems~\cite{10591448}.
We further allow \textit{directionality}, which reflects the asymmetric nature of influence in online platforms (e.g., broadcaster-listener dynamics) and is compatible with extensions where influence weights may evolve and take both positive and negative values~\cite{disaro2024homophily}.
Within this setting, we derive theoretical guarantees showing when equilibrium disruption can increase and how to perturb innate opinions to maximally amplify disruption~\cite{chen2020networkdisruptionmaximizingdisagreement}. 

We operationalize these results by proposing an empirical framework based on Large Language Models (LLMs). Following~\cite{coppolillo2025engagementdrivencontentgenerationlarge}, we use reinforcement learning to fine-tune an LLM to generate targeted, disruptive messages. {This yields an end-to-end pipeline that links (i) a network-level disruptive objective, (ii) an opinion-dynamics model, and (iii) a text generation policy.}
Experiments on synthetic graphs and real-world social graphs from $\mathbb{X}$ show that the fine-tuned LLM can inject content that maximizes dissensus, with the resulting disruption approaching the theoretical upper bounds predicted by our analysis.

\noindent{\textbf{Summary.}} Our {contributions} can be hence summarized as follows:
\begin{itemize}[leftmargin=*]
\item We formalize \textbf{disruption} as an optimization objective and show that the {basic} FJ model is structurally inadequate for increasing disruption at equilibrium.
\item We study an \textbf{extended} FJ model with heterogeneous susceptibility and negative (directional) influence, and provide theoretical guarantees for \textbf{maximally perturbing} innate opinions to amplify disruption.
\item We introduce a reinforcement-learning framework to \textbf{fine-tune} LLMs for disruption-oriented content generation, grounding the training signal in the above theory.
\item We \textbf{validate} the approach on synthetic and real-world networks, showing that generated content can approach the theoretical disruption limits.
\end{itemize}
\section{Related Work}
\label{sec:related}
\paragraph{Opinion Dynamics Models} Opinion dynamics models describe how individual beliefs evolve under social influence. The DeGroot model~\cite{DING201962} assumes agents update opinions by averaging their own with neighbors, driving consensus even under strong homophily. However, its high sensitivity to noise can lead to indefinite opinion drift and systemic loss~\cite{banerjee2023consensusdisagreementinformationaggregation}. The Friedkin-Johnsen (FJ) model~\cite{friedkin1990social} generalizes DeGroot by incorporating \textit{innate opinions}, agents intrinsic adherence to initial beliefs, into each update. This anchoring improves robustness, avoiding the unbounded drift of DeGroot~\cite{banerjee2023consensusdisagreementinformationaggregation}. The FJ model has been widely validated in the literature, with extensions for memory and multi-hop influence~\cite{fj-mm}, multidimensional and private/public opinions~\cite{friedkin1999social}, and susceptibility-weighted updates~\cite{ghaderi2014opinion}. 
A closely related line of work studies consensus networks with \emph{antagonistic (signed) interactions}, showing that negative ties fundamentally change the asymptotic behavior: instead of standard consensus, trajectories may converge to \emph{bipartite consensus} under structural-balance conditions~\cite{altafini2013consensus}, and more generally can exhibit persistent polarization into hostile camps depending on the signed topology~\cite{proskurnikov2016opinion}. These results provide a dynamical-systems foundation for modeling disagreement induced by antagonism.
In the context of network disruption, simple FJ variants \textit{prevent} higher disruption at equilibrium compared to the initial state~\cite{biondiboldrini}, limiting the model to represent real-world settings where polarization and disagreement increase over time~\cite{bail2018exposure, cinelli2021echo, cinelli2021echo}. To overcome such limitation, we build our disruptive framework on a generalized FJ model that integrates (i) varying susceptibility to persuasion~\cite{10.1145/3219819.3219983}, (ii) negative influence among users (signed interactions)~\cite{10591448}, and (iii) crucially allowing \textit{directionality}. Notably, our focus is not on characterizing when polarization \emph{emerges}, but on how it can be \emph{strategically amplified} via targeted content injection. 

\vspace{-0.13cm}
\paragraph{Polarization and Disagreement} Quantifying polarization and disagreement has been central to computational social science. Common measures include opinion variance, neighbor disagreement~\cite{minimizing-polarization}, extremeness indices~\cite{biondi2023measuring}, and community boundaries~\cite{Guerra_Meira}. We adopt variance for polarization and edge-level divergence for disagreement~\cite{minimizing-polarization}. While most work seeks to \textit{minimize} polarization and disagreement~\cite{minimizing-polarization, NEURIPS2021_101951fe, 10.1145/3583780.3615025}, some explore disruption through targeted opinion manipulation~\cite{chen2020networkdisruptionmaximizingdisagreement, 10831299}. In influence maximization~\cite{kempe2003maximizing}, related efforts address misinformation containment and adversarial influence~\cite{gionis2013opinion, goldman2021maximizing}. Goldman et al.~\cite{goldman2021maximizing} model disruption via FJ by manipulating node opinions to maximize disagreement, but compare only equilibrium states after manipulation. In contrast, we provide theoretical guarantees for perturbing innate opinions to maximize equilibrium disruption from the \textit{initial} state. Gaitonde et al~\cite{gaitonde2020adversarialperturbationsopiniondynamics} resorts to spectral techniques to amplify disagreement into the network, however assuming graphs to be undirected.  

\paragraph{LLMs in Social Influence} LLMs have transformed content generation, enabling automated, personalized, and optimized messaging across platforms~\cite{10.1145/3706599.3720019, griffin-etal-2023-large, coppolillo2025engagementdrivencontentgenerationlarge, personalizedemail, 10654534}. Integrated with opinion dynamics, LLMs shift from passive simulation to active intervention, becoming agents capable of influencing social outcomes. Recent studies show LLMs can match or exceed human persuasion in various contexts~\cite{rogiers2024persuasionlargelanguagemodels, Salvi_2025, carrascofarre2024largelanguagemodelspersuasive}. Reinforcement learning (RL) is widely used for LLM alignment~\cite{ouyang2022training} and has been applied to persuasion~\cite{hiraoka2014reinforcement} and negotiation~\cite{lewis2017deal}. In social influence, Coppolillo et al.~\cite{coppolillo2025engagementdrivencontentgenerationlarge} fine-tune LLMs with reward functions to steer public discourse. Building on this, we use RL to induce LLMs to disrupt social graphs, illustrating how such manipulation could be operationalized. This raises significant ethical concerns about large-scale opinion manipulation and potential impacts on political processes~\cite{scalingLLMs}.
\section{Preliminaries and Context}
\label{sec:model}


We start by establishing the notation that will be used throughout this section. Vectors and matrices are written in boldface. For a vector $\mathbf{x}$, its $i$-th component is denoted by $x_i$. Furthermore, let $\mathbf{1}$ denote the vector whose entries are all equal to $1$, and let $\mathbf{e}_i$ denote the vector whose entries are all $0$ except for the $i$-th entry, which is $1$. The identity matrix is denoted as $I$. 

Let $\graph = (\nodes, \edges)$ be an undirected social graph where \nodes are users and \edges are connections among them, with $\nnodes = |\nodes|$ and $\nedges = |\edges|$. Let $A$ represent the adjacency matrix of \graph, with  $d_i = \sum_{j} a_{i,j}$ denoting the degree of node $i$ and the matrix $D = \mathit{diag}(d_1, \ldots d_n)$ the corresponding degree matrix. 

Let \innateopinions $\in [-1, 1]^{\nnodes}$ denote the \textit{innate} users opinion vector concerning a given topic. The Friedkin-Johnson model~\cite{friedkin1990social} defines the temporal evolution of a vector \expressedopinions $\in [-1, 1]^\nnodes$ of  \textit{expressed} opinions, computed according to the equations:
\begin{equation}
\begin{split}
    \expressedopinion_i(0) = &\ s_i\\
    \expressedopinion_i(t+1) = &\ \frac{s_i +  \sum_{j: a_{i,j} = 1} \expressedopinion_j(t)}{1 + d_i}    
\end{split}
\end{equation}
The above recursive equation admits a solution as $\expressedopinions^*$, the opinion vector at equilibrium, which can be computed as
\begin{equation}\label{eq:equilibrium}
    \expressedopinions^* = (I + L)^{-1} \innateopinions
\end{equation}
Here, $L$ is the combinatorial Laplacian of  \graph, defined as $L = D - A$. This characterization is summarized in Appendix~\ref{eq:appendix-fj-model}. 

\noindent{\textbf{Disruption.}} Given a generic opinion vector $\genericvec \in [-1,1]^n$, we define the \textit{disruption} $\disruption_{\graph, \genericvec}$ of a social graph \graph conditioned by \genericvec as in~\cite{minimizing-polarization}:
\begin{equation}
    \disruption_{\graph,\genericvec} = P_{\graph,\genericvec} + D_{\graph,\genericvec}
\label{eq:disruption}
\end{equation}
where $P_{\graph,\genericvec}$ quantifies the polarization of \graph while $D_{\graph,\genericvec}$ represents its disagreement. The two measures are defined as follows.

\noindent{\textbf{Polarization.}} Following the standard definition of variance, let $\overline{\genericvec}$ be the mean-centered equilibrium vector:
\begin{equation}
    \overline{\genericvec} = \genericvec - \frac{{\genericvec}^\top\mathbf{1}}{\nnodes}\mathbf{1}
\end{equation}
%
Then the polarization  is defined to be:
\begin{equation}
    P_{\graph,\genericvec} = \sum_{\node \in \nodes} \overline{\genericvec}^2_u = \overline{\genericvec}^\top\overline{\genericvec}
\label{eq:polarization}
\end{equation}



\noindent{\textbf{Disagreement.}} Given an edge $(u,v)$, let the disagreement $d(u,v)$ be the squared difference between the opinions of $u, v$ at equilibrium:
\begin{equation}
  d_{\genericvec}(u,v) = (x_u - x_v)^2
\end{equation}
The overall disagreement $D_{\graph,\genericvec}$ is hence defined as:
\begin{equation}
D_{\graph,\genericvec} = \sum_{(u,v) \in \edges} d_{\genericvec}(u,v).
\label{eq:disagreement}
\end{equation}

We can rewrite the disruption score $\disruption_{\graph, \genericvec}$ as follows:
\begin{equation}
\begin{aligned}
\disruption_{\graph, \genericvec} &= {P}_{\graph,\genericvec} + {D}_{\graph,\genericvec} = \genericvec^\top (I + L - \frac{1}{n}\mathbf{1}\mathbf{1}^\top) \genericvec \\
\end{aligned}
\label{eq:rewritten-distruption}
\end{equation}


The first aspect we study is whether the underlying FJ model can disrupt the network. In mathematical terms, this can be translated in the following property: 
\begin{equation}
\disruption_{\graph,\expressedopinions^*} > \disruption_{\graph,\innateopinions}
\end{equation}
meaning that the disruption computed on the graph at equilibrium is higher than the one obtained on the initial network.

Let $H  =  I + L - \frac{1}{n}\mathbf{1}\mathbf{1}^\top$. 
By combining Equations~\ref{eq:equilibrium} and~\ref{eq:rewritten-distruption}, we can rewrite:
\begin{equation}
\begin{aligned}   
\disruption_{\graph, \expressedopinions^*} &= P_{\graph,\expressedopinions^*} + D_{\graph,\expressedopinions^*} = {\expressedopinions^*}^\top H \expressedopinions^* \\
     &= {\innateopinions}^\top (I + L)^{-1} H (I + L)^{-1} \innateopinions
\end{aligned}
\end{equation}
and 
\begin{equation}
\begin{aligned}   
\disruption_{\graph, \innateopinions} &= P_{\graph,\innateopinions} + D_{\graph,\innateopinions} = {\innateopinions}^\top H\innateopinions \\
\end{aligned}
\end{equation}




\begin{restatable}{resth}{keyLemma}
$\disruption_{\graph, \expressedopinions^*}\leq \disruption_{\graph, \innateopinions}$.  
\label{th:semi-definite}
\end{restatable}
The proof is provided in Appendix~\ref{sec:proof}. 
This result shows that the basic formulations of the FJ model \emph{cannot induce network disruption at equilibrium}. As aforementioned, the reason is structural: the classical FJ framework assumes that social interactions are purely assimilative, i.e., every neighbor exerts a non-negative influence on an individual expressed opinion. As a consequence, equilibrium opinions remain confined to the convex hull of the innate opinions, preventing the emergence or amplification of disagreement or polarization.

Such behavior is increasingly misaligned with empirical evidence from online social systems, where interactions are often antagonistic, asymmetric, and explicitly conflictual. Real-world opinion dynamics routinely exhibit repulsion, backlash effects, echo chambers, and persistent or growing polarization~\cite{garimella2018quantifying, cinelli2021echo, morales2015measuring, bail2018exposure, conover2011political, del2016echo}. These phenomena cannot be reliably captured by a model in which influence is assumed to be uniformly positive and symmetric. In this sense, the standard FJ model provides a mathematically convenient but \emph{behaviorally unrealistic} abstraction of social influence.

To address this mismatch, prior work~\cite{10.1145/3219819.3219983,10591448} has extended the FJ model by introducing susceptibility and signed influence on undirected graphs, allowing edges to encode antagonism in addition to attraction. While such formulations extensions already move beyond purely positive, homogeneous influence, {they still assume undirected (symmetric) interactions. This distinction is not merely representational: directionality changes how antagonism and susceptibility propagate through the network, and it is essential to model typical influence patterns and opinion dynamics. }
\noindent\textbf{Directionality.} 
We overcome this limitation by adopting a strictly more general formulation that combines heterogeneous susceptibility with \emph{directional} influence and antagonism. Let $W$ be a weighted influence matrix where each entry $w_{ij} \in \{-1,0,1\}$ captures the (possibly negative) effect exerted by user $j$ on user $i$, which is not necessarily reciprocal. This directionality reflects a key empirical property of modern social platforms, where influence is inherently asymmetric. This is what typically can be observed between content creators and consumers, highly connected hubs and peripheral users, or authoritative and reactive accounts.


Formally, for each node $i$, we define $d_i = \sum_j |w_{ij}|$ and set $D=\mathit{diag}(d_1, \ldots, d_n)$. 
Further, let $\lambda_i$ denote the susceptibility of user $i$, and $\Lambda = \mathit{diag}(\lambda_1,\ldots,\lambda_n)$. The resulting update rule is
\begin{equation}
\expressedopinion_i(t+1) = \ (1 - \lambda_i) \innateopinion_i + \frac{\lambda_i}{d_i} \sum_{j} w_{ij} \expressedopinion_j(t)    
\end{equation}
In Appendix~\ref{sec:proof}, we establish theoretical guarantees on matrix invertibility and spectral properties, showing that the model admits a unique equilibrium under mild conditions. The closed-form expression for the equilibrium expressed-opinion vector is given by
\begin{equation}
    \begin{split}
    \expressedopinions^{*} = 
    (I - \Lambda D^{-1}W)^{-1} (I - \Lambda)
    \label{eq:z-equilibrium-extended}
    \end{split}
\end{equation}

\begin{figure}
    \centering
    \begin{subfigure}[b]{0.57\columnwidth}
        \includegraphics[width=\columnwidth]{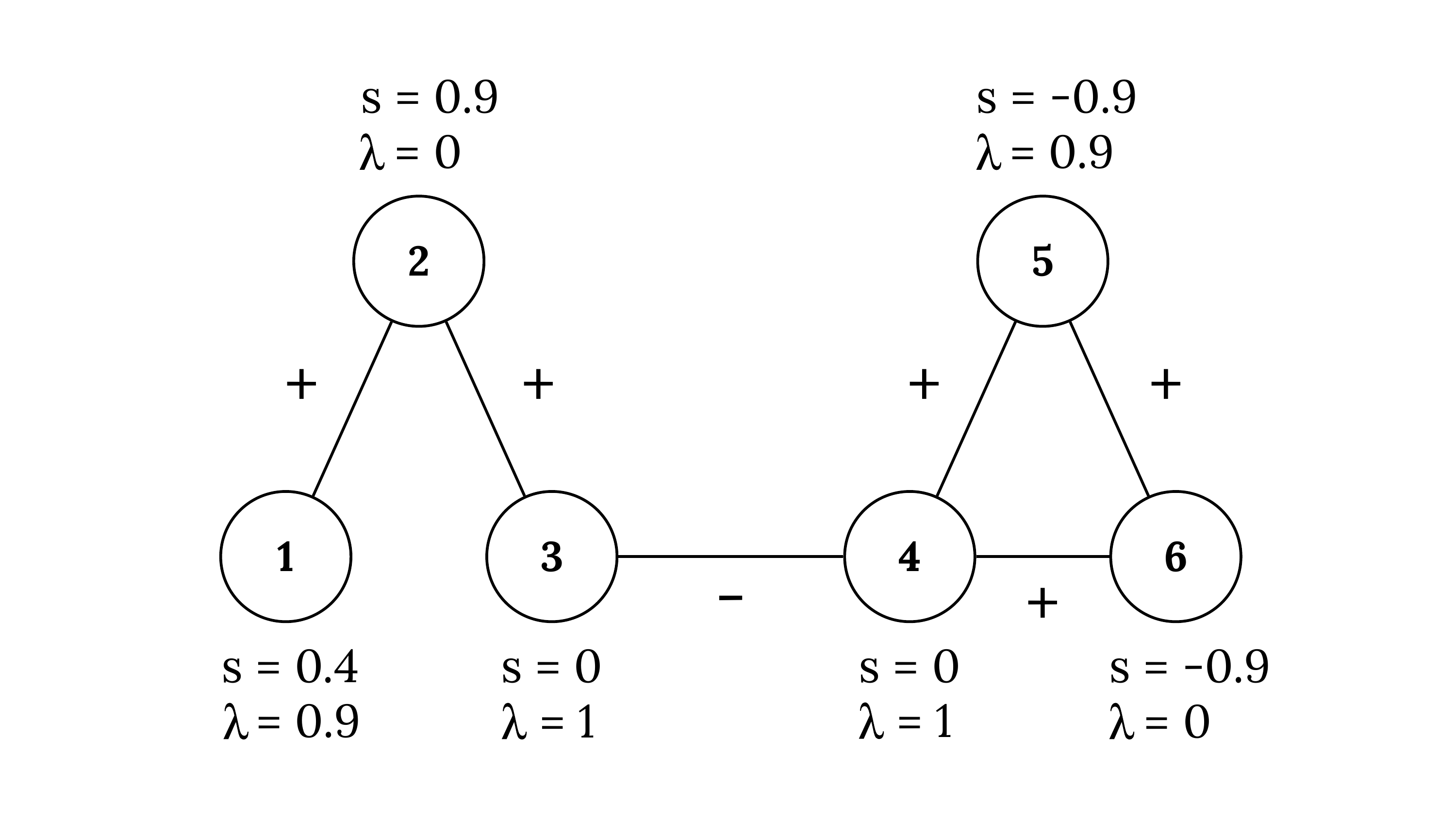}
        \caption{}
    \end{subfigure}  
    \begin{subfigure}[b]{0.42\columnwidth}
        \includegraphics[width=\columnwidth]{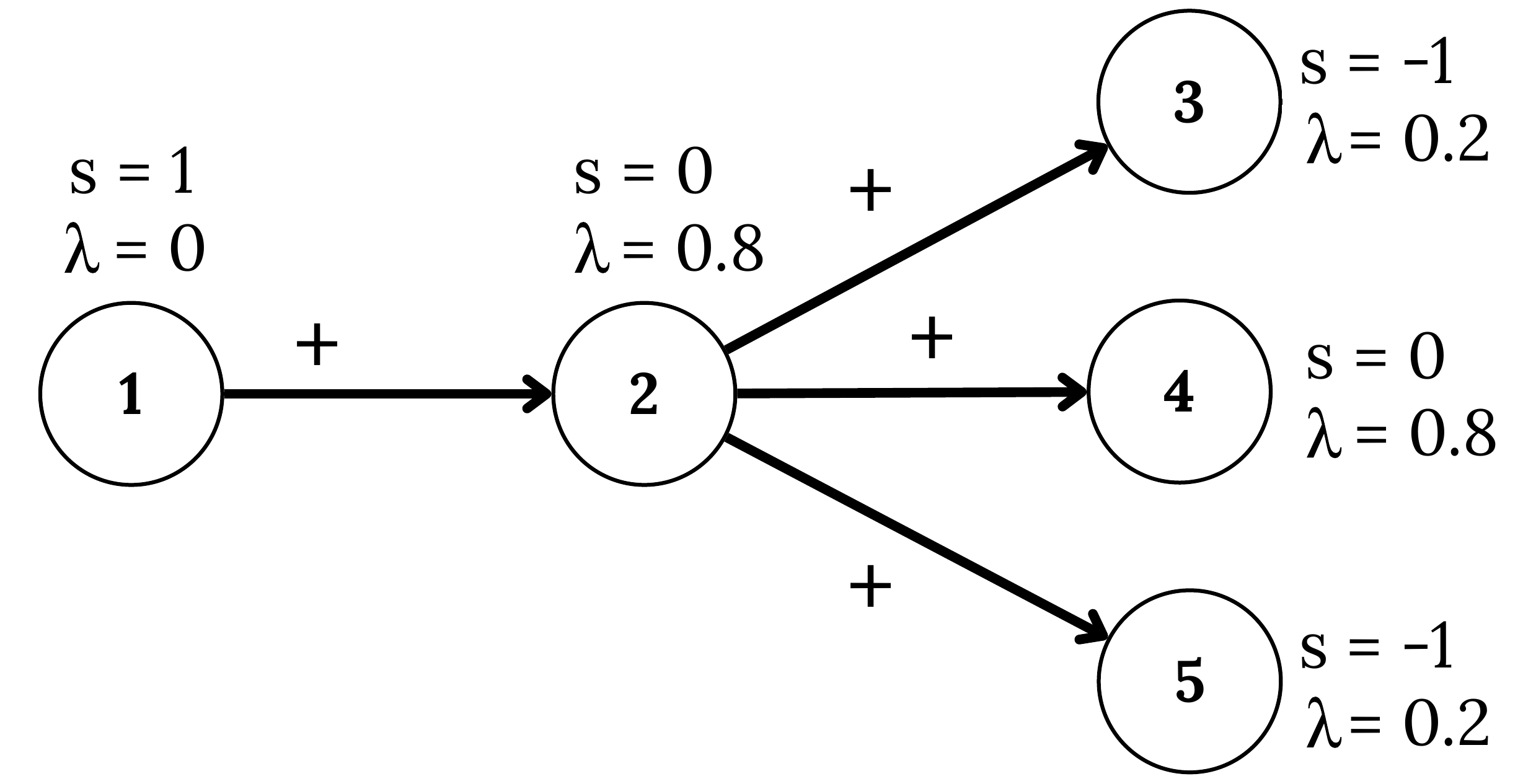}
        \caption{}
    \end{subfigure}  
    \caption{Examples of graphs with symmetric (left) and asymmetric (right) influence matrices, empirically showing that $\disruption_{\graph,\expressedopinions^*} > \disruption_{\graph, \innateopinions}$. Here, the values of \textbf{s} and $\lambda$ denote the innate opinions and susceptibilities of the nodes, respectively, while the edges indicate node influence ($W$).}
    \label{fig:graph-example}
\end{figure}

{Let $H = (I - \Lambda D^{-1}W)^{-1} (I - \Lambda)$. The disruption scores for this model are therefore defined as:
\begin{equation}
\disruption_{\graph, \expressedopinions^*} = \innateopinions^\top M^\top H M \innateopinions \quad \mathrm{and} \quad \disruption_{\graph, \innateopinions} = \innateopinions^\top H \innateopinions.
\end{equation}
}
Again, by analyzing $\disruption_{\graph, \expressedopinions^*} - \disruption_{\graph, \innateopinions}$, we can conclude that the addition of antagonism and directional influence can actually induce disruption. 
%
%
To illustrate this,
we consider the two graphs $\graph^\prime$, $\graph^{\prime\prime}$ depicted in Figure~\ref{fig:graph-example}. 
%
%
For both configurations, the equilibrium disruption strictly exceeds the disruption at the initial state, namely:
\begin{equation*}
    \disruption_{\graph^{\prime}, \expressedopinions^{\prime*}} = 8.01, \quad \text{and} \quad \disruption_{\graph^{\prime}, \innateopinions^\prime} = 5.23,
\end{equation*}
\begin{equation*}
    \disruption_{\graph^{\prime\prime}, \expressedopinions^{\prime\prime*}} = 10.79, \quad \text{and} \quad \disruption_{\graph^{\prime\prime}, \innateopinions^{\prime\prime}} = 10.0
\end{equation*}

The examples demonstrate that disruption can increase at equilibrium both in the presence of symmetric antagonistic interactions (as in $\graph^\prime$) and and under purely asymmetric influence (as in $\graph^{\prime\prime}$).
%
%
In $\graph^\prime$, two homophilic communities are connected through a small set of hubs with \emph{opposite} influence. Thus, within-community attraction contracts opinions, while antagonistic cross-community influence increases inter-community distances, yielding higher equilibrium disagreement and hence larger $\disruption_{\graph^\prime, \expressedopinions^{*\prime}}$, a pattern consistent with echo-chamber formation observed in real online debates~\cite{cinelli2021echo}.
%
Graph $\graph^{\prime\prime}$ instead highlights the role of \emph{directionality}. A hub (node $2$) influences many downstream nodes while being influenced by a smaller upstream neighborhood; even without negative ties, this asymmetric flow can shift the hub (and its downstream neighborhood) in a way that widens gaps with less-influenced (or more stubborn) nodes, increasing $\disruption_{\graph^{\prime\prime},\expressedopinions^{*\prime\prime}}$. 
Notably, if we discard directionality and consider the undirected variant of $\graph^{\prime\prime}$, namely $\tilde{\graph}^{\prime\prime}$, the resulting disruption score is $\disruption_{\tilde{\graph}^{\prime\prime}, \expressedopinions^{\prime\prime*}} = 7.48$. This suggests that directionality has a crucial role in shaping the disruption of the underlying graph.

The above examples are not intended as exhaustive empirical validations, but rather as minimal, transparent configurations that isolate the mechanisms through which disruption can increase at equilibrium. More broadly, disruption can emerge in extended frameworks that incorporate antagonism and/or directional influence. This naturally raises the question of how such disruption can be deliberately amplified: given a social network and the proposed dynamics, \textit{which minimal intervention on an agent innate opinion maximizes the resulting equilibrium disruption?}
\color{black}
We formalize this in the following problem formulation:
\begin{tcolorbox}[colback=gray!10, colframe=black, rounded corners]
\begin{problem}
    Given a social network \graph, the inner opinion vector \innateopinions, and a source node \node, we want to find a variation of $\innateopinion_\node$ to maximally disrupt \graph at equilibrium. That is, we aim to find the value $\alpha$ which modifies \innateopinions such that 
    \begin{equation}
        \innateopinions'= \innateopinions - \alpha \mathbf{e}_\node
    \label{eq:s_first}
    \end{equation} maximizes $\disruption_{\graph, \expressedopinions^*}$, with $\innateopinion_u - 1 \leq \alpha \leq \innateopinion_\node + 1$.
\end{problem}
\end{tcolorbox}

Let $B = M^\top H M$. By using Equations~\ref{eq:disruption} and~\ref{eq:z-equilibrium-extended}, our problem can be formalized as:


\begin{equation}
\max_{\innateopinion_u - 1 \leq \alpha \leq \innateopinion_\node + 1} \quad (\innateopinions - \alpha \mathbf{e}_\node)^\top B (\innateopinions - \alpha \mathbf{e}_\node) 
\label{eq:problem-formulation}
\end{equation}
Notably, this problem admits a remarkably simple and structural solution stating that the optimal intervention always lies at the extremes of the opinion domain.

\begin{restatable}{resth}{MaximizationLemma}\label{theo:maximization}
    The value of $\alpha$ that maximizes Equation~\ref{eq:problem-formulation} is always equal to either of its bounds.
\end{restatable}

The proof is detailed in the Appendix~\ref{sec:proof}. Notably, our result is consistent with previous work~\cite{chen2020networkdisruptionmaximizingdisagreement}, that however assumes networks to be undirected and unsigned.


\noindent{\textbf{Modeling Assumptions.}} We emphasize a crucial modeling choice that distinguishes our approach from prior work. Although similar problems have been studied where interventions target multiple peers simultaneously~\cite{10.1145/3580305.3599255, gionis2013opinion, masuda2015opinion,Sun_Zhang_2023}, we argue that such settings are unrealistic in the context of adversarial content manipulation on social networks. In real-world scenarios, an adversarial agent (e.g., an automated bot, a coordinated inauthentic account, or a malicious language model) would not have direct access to the internal opinion states of arbitrary users across the network, nor the capability to directly perturb the beliefs of multiple nodes. Such assumptions would require either platform-level access (as in~\cite{musco2018minimizing}) or coordinated manipulation of many accounts, which faces practical barriers including detection mechanisms~\cite{ferrara2016rise, cresci2017paradigm} and resource constraints.
We instead adopt a more realistic threat model where the adversary operates as a \textit{single} node within the network it aims to disrupt. The adversarial agent can only control its own expressed opinion state ($s_u$) and influence others through the standard mechanisms of social interaction, such as posting content, engaging in discussions, and participating in the network's information diffusion processes. This constraint aligns with documented cases of individual influential actors or sophisticated bot accounts that leverage strategic positioning and content optimization to amplify polarization~\cite{bail2018exposure, tucker2018social}. 

Our setting is therefore more conservative yet practically grounded: rather than assuming omniscient manipulation capabilities, we model an adversary that must work within the attacker constrained by the network actual structural and informational constraints of the network, which makes our findings more readily relevant for real-world mitigation strategies. {Nevertheless, Theorem~\ref{theo:maximization} naturally generalizes to the case of $k$ known peers under adversarial control. In other words, if the attacker can directly manipulate a predefined set of nodes (for instance, bots injected in the network), the corresponding optimization task similarly amounts to determining how to adjust the opinions of these nodes so as to maximize the induced disruption. Importantly, this extended optimization problem can be solved by pushing each controlled node to one of its boundaries (see Appendix~\ref{sec:proof}).}

\section{Disruption Framework}
\label{sec:empirical-framework}
Once proved how to manipulate the original innate opinions to induce maximum disruption on the social graph, we consider a scenario where the opinion of a user is represented by a piece of textual content (e.g., post, comment, etc.) they share on the network. Motivated by this assumption, we tackle the last research objective: automatically obtain textual content which aligns with the manipulated opinion $\innateopinion'_u$. To do this, we rely on an empirical framework: exploiting a Large Language Model specifically fine-tuned for achieving this disruptive goal.

In the following, we discuss the main components of such a disruptive framework: Large Language Models and the Fine-tuning procedure based on Reinforcement Learning.

\paragraph{Large Language Models}
A Large Language Model (LLM) can be formally characterized as a stochastic function $\mathrm{LLM}_{\theta}(x) = y$, which maps an input token sequence $x = [x_1, x_2, \dots, x_k]$ to an output sequence $y = [y_1, y_2, \dots, y_l]$, 
where $k$ and $l$ denote the lengths of the input and output, respectively. The model defines a conditional probability distribution $P_\theta(y|x)$ over possible outputs given the input, capturing intricate dependencies and semantic structures in natural language. A response $y$ is then sampled from this distribution, i.e., $y \sim P_\theta(\cdot|x)$.

In our approach, we fine-tune the LLM using a Reinforcement Learning (RL) framework, a widely adopted methodology for optimizing LLMs~\cite{10.5555/3600270.3602281}. RL is particularly well-suited for settings where an agent must learn to operate within a dynamic environment by refining its behavior based on feedback, typically in the form of rewards or penalties. The agent objective is to develop a \textit{policy}, i.e., a strategy guiding its actions, that maximizes cumulative reward. Notably, although less computationally demanding baselines could be considered (e.g., prompting), prior work~\cite{coppolillo2025engagementdrivencontentgenerationlarge} has established that RL is considerably more effective in information diffusion scenarios. This is particularly critical for negative-oriented target content, as baseline strategies tend to reflect the LLM intrinsic safeguards.

Within our framework, given a prompt $x$, we sample a response $y \sim P_\theta(\cdot|x)$ and evaluate it using a task-specific reward function $\mathcal{R}(y)$. This reward guides optimization via a policy gradient method, which aims to maximize the following objective:
\begin{equation}
    \mathcal{L}(\theta) = \mathbb{E}_{y\sim P_\theta(\cdot|x)} \left[ \mathcal{R}(y) - \beta \log \frac{P_\theta(y|x)}{P_{\theta'}(y|x)} \right],
\end{equation}

where $\theta'$ corresponds to the parameters of a reference (pre-trained but not fine-tuned) model. The second term represents a KL-divergence regularization between the fine-tuned and the reference model, aimed to prevent output drift. The scalar $\beta$ modulates this penalty and is set to the default value of 0.05 in our experiments.\footnote{\url{https://github.com/huggingface/trl/blob/main/trl/trainer/ppo_config.py}}

To enhance the stability of the learning process, refinements to the above formulation have been proposed. In particular, we adopt the Proximal Policy Optimization (PPO) algorithm~\cite{schulman2017proximal}, which mitigates instability by restricting the magnitude of policy updates through a clipped objective function. This constraint helps maintain reliable and efficient learning, ensuring that the policy evolves in a controlled manner during fine-tuning.

\paragraph{Fine-tuning Procedure} 
Fixed a source node \node, to produce a textual content whose opinion aligns with $\innateopinion'_{\node}$, i.e., the value which disrupts the network \graph (Equation~\ref{eq:problem-formulation}), we develop a fine-tuning framework based on RL, consisting of the following steps:

\begin{enumerate}[leftmargin=*]
    \item We prompt the LLM with a message \prompt to generate a content \content about the topic: $c = \textrm{LLM}_{\theta}(q).$
    In our experiments, $\prompt =$ ``{Generate a post about} \texttt{[TOPIC]}.'' In this step, we also compute the KL-divergence between the reference and the fine-tuning LLM.
    \item We use a function $\mathcal{S}$ to compute the opinion of the generated content $c$, denoting it as $x$, i.e., $x = \mathcal{S}(c)$. The implementation details of $\mathcal{S}$ is provided in the Settings paragraph below.
    
    \item We define the reward $\mathcal{R}$ as:
    \begin{equation}
        \mathcal{R}(x) = e^{-\frac{(x - s'_u)^2}{2\sigma^2}}
    \label{eq:reward}
    \end{equation}
    implying that $\mathcal{R}(x)$ decays exponentially when $x$ departs from $\innateopinion'_\node$. In our experimental setting, we set $\sigma = 0.15$. 
    \item We use $\mathcal{R}$ to update the PPO policy, and repeat the process. 
\end{enumerate}
The training procedure stops if one of the following conditions occurs: (\textit{i}) a content with opinion $x \approx \innateopinion'_\node$ is generated, or (\textit{ii}) the KL-divergence $\kappa$ exceeds a fixed threshold $\tau$, or (\textit{iii}) the number of iterations approaches a maximum value $T$. The content opinion $x$ approaches $\innateopinion'_\node$ if $x \in [\innateopinion'_\node - \epsilon, \innateopinion'_\node + \epsilon]$. In our experiments, we set $\epsilon = 0.05$.
If convergence is reached, i.e., if (\textit{i}) occurred, given the theoretical guarantees provided in Section~\ref{sec:model}, then the LLM generated a content \content which induces the maximum disruption over the network. We denote such content as $\content^*$.
A sketch of the overall training procedure is depicted in Algorithm~\ref{alg:procedure} in the Appendix.
\begin{figure}[!ht]
    \centering
    \includegraphics[width=0.4\linewidth]{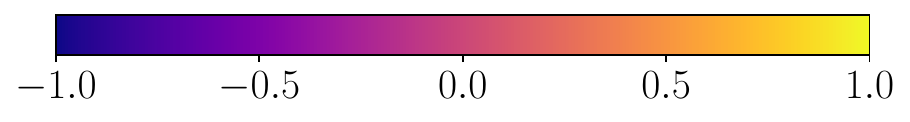}\\
    \begin{subfigure}[b]{0.23\linewidth}
        \includegraphics[width=\linewidth]{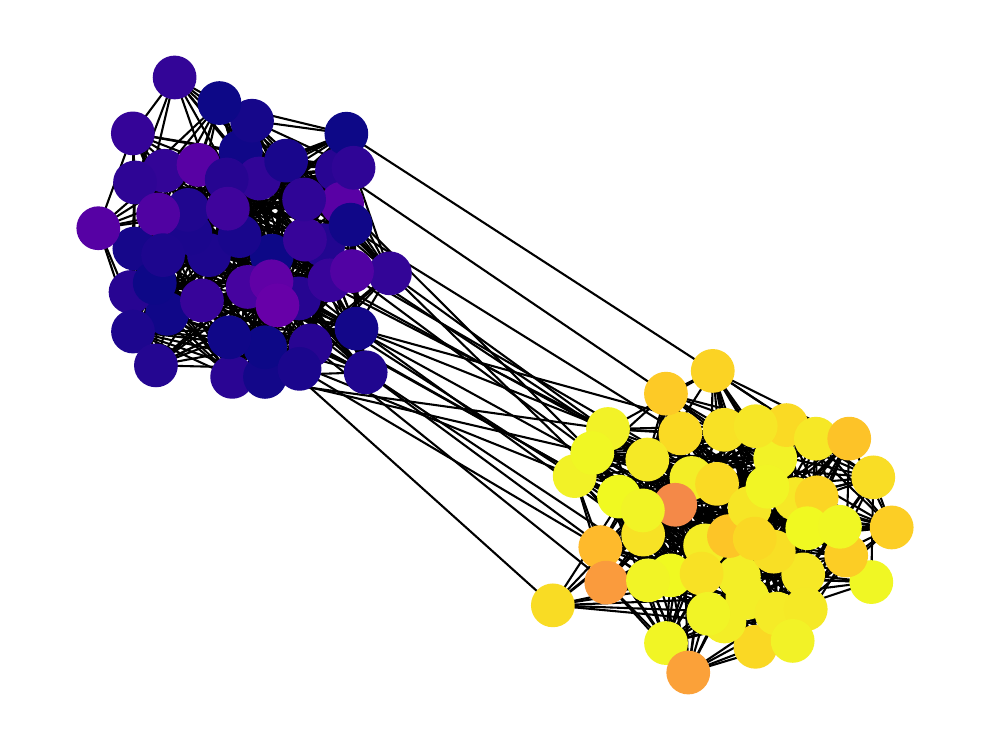}
        \caption{$\alpha_1 = \beta_2= 1$}
        \label{}
    \end{subfigure}
\begin{subfigure}[b]{0.23\linewidth}
\includegraphics[width=\linewidth]{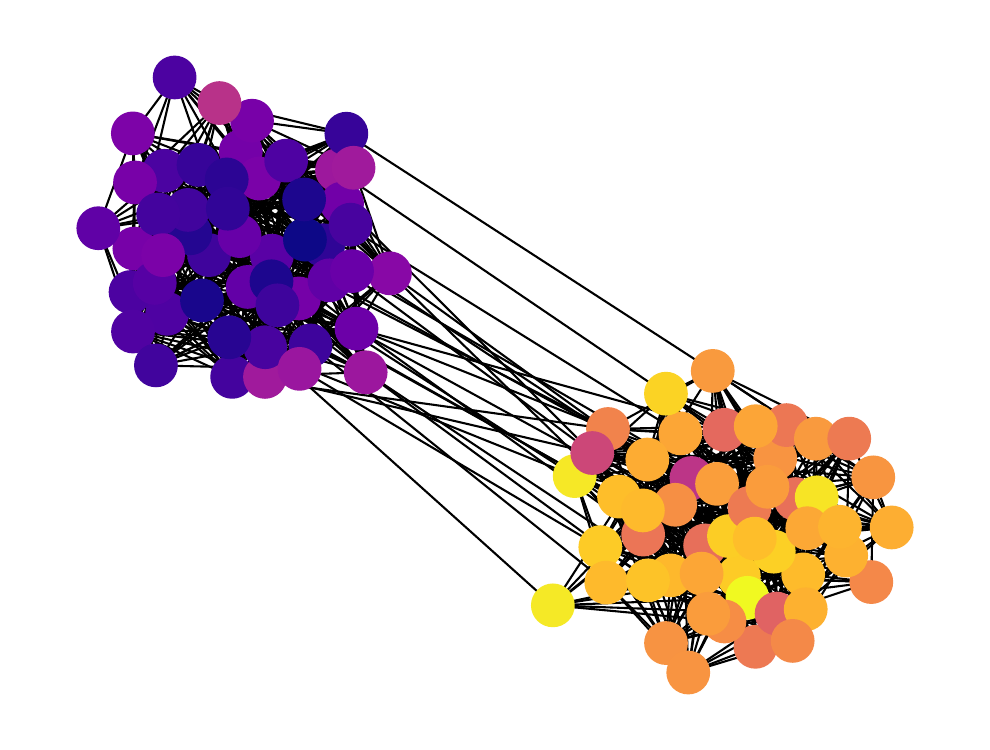}
\caption{$\alpha_1=\beta_2=5$}
        \label{}
\end{subfigure}
\begin{subfigure}[b]{0.23\linewidth}
\includegraphics[width=\linewidth]{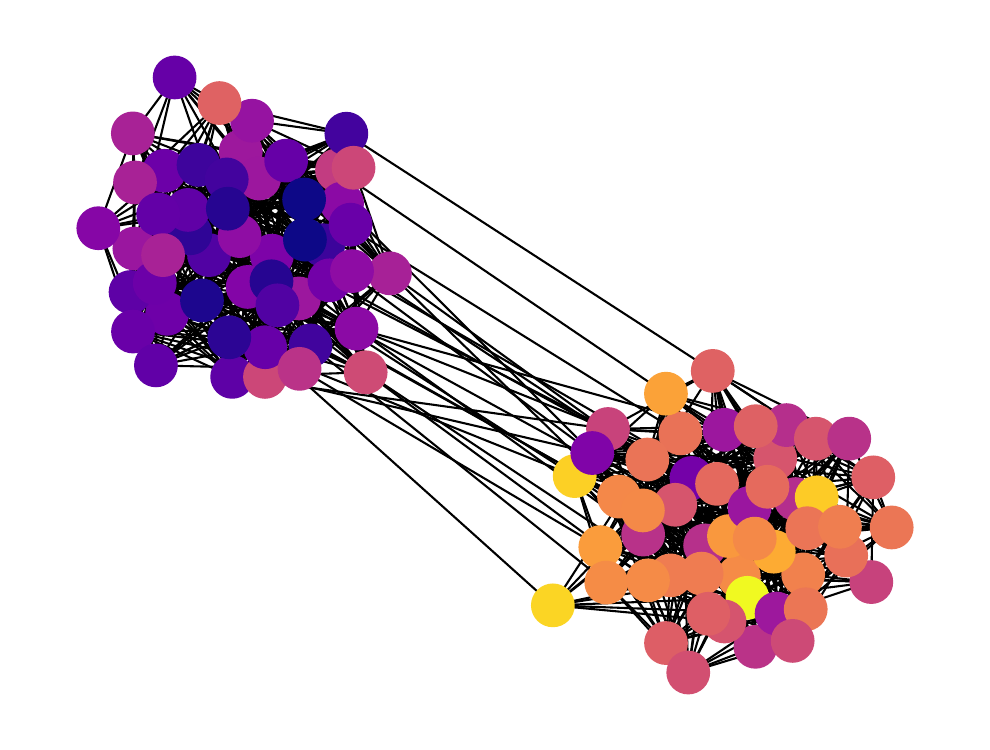}
\caption{$\alpha_1=\beta_2=10$}
        \label{}
\end{subfigure}
\begin{subfigure}[b]{.23\linewidth}
\includegraphics[width=\linewidth]{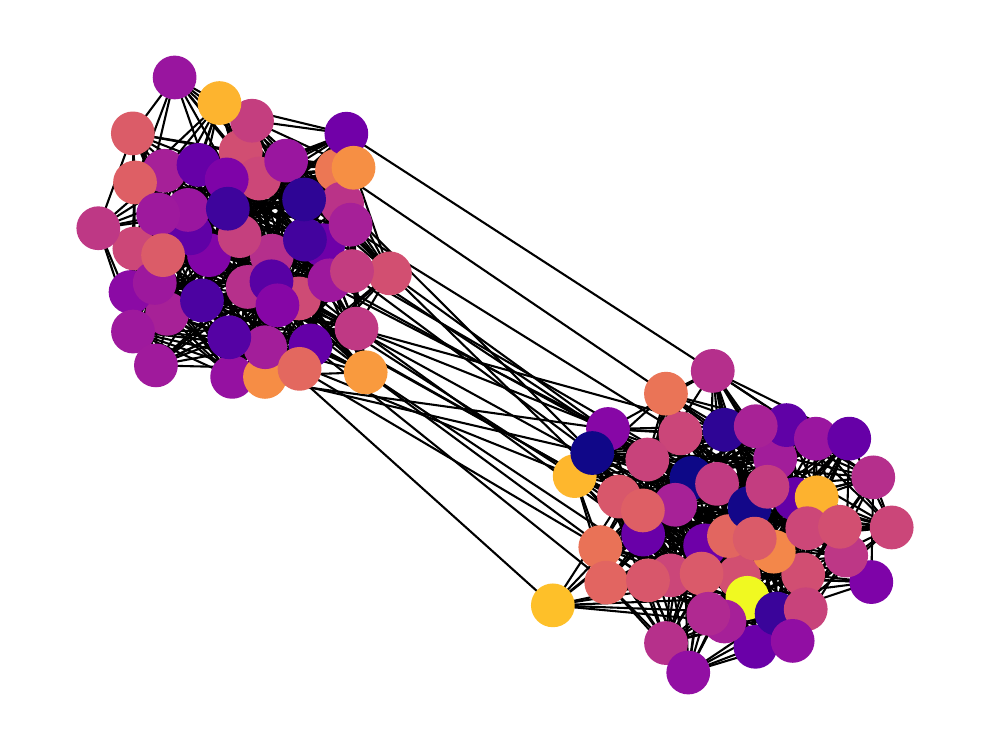}
\caption{$\alpha_1=\beta_2=15$}
        \label{}
\end{subfigure}
    \caption{Opinion distribution of the synthetic graph by varying the Beta parameters $\alpha_1$ and $\beta_2$, from high (left) to low homophily (right). We fix $\alpha_2 = \beta_1 = 15$ for all configurations.}
    \label{fig:synthetic-networks-opinions}
\end{figure}

\begin{figure*}[!ht]
    \centering

    \includegraphics[width=0.2\linewidth]{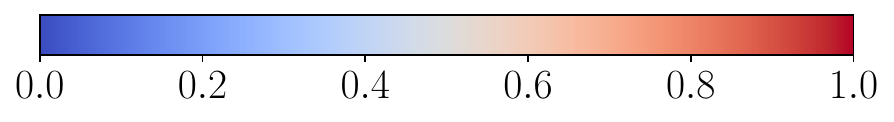}\\
\begin{subfigure}[b]{0.13\linewidth}
    \includegraphics[width=\linewidth]{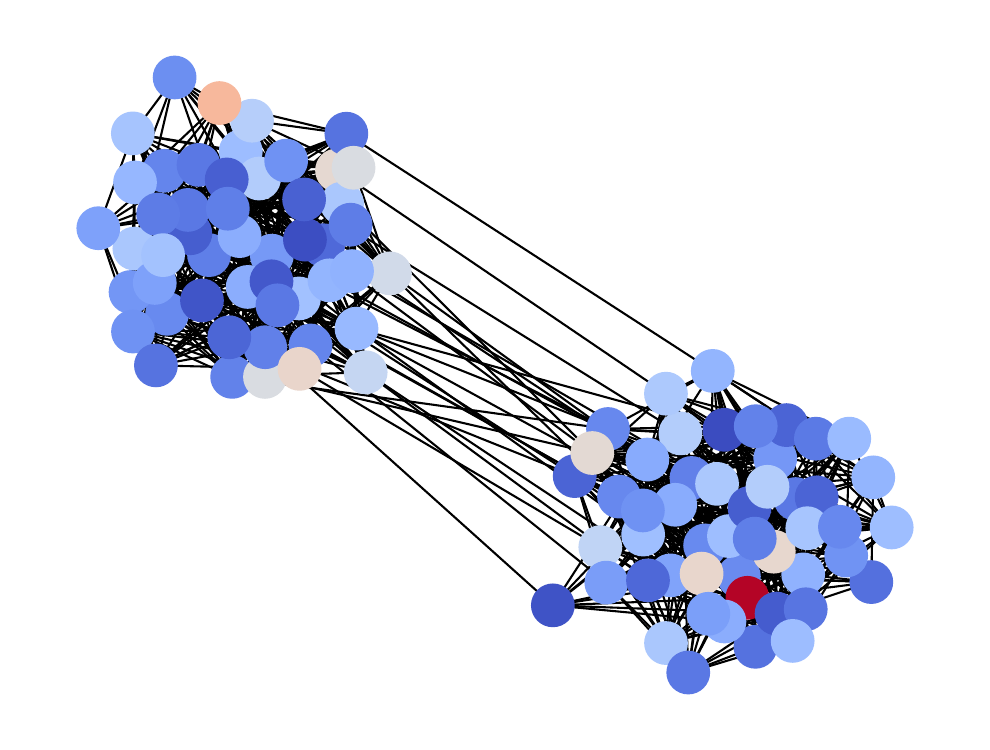}
    \caption{$\alpha_\Lambda=2, \beta_\Lambda=15$}
\end{subfigure}
\begin{subfigure}[b]{0.13\linewidth}
\includegraphics[width=\linewidth]{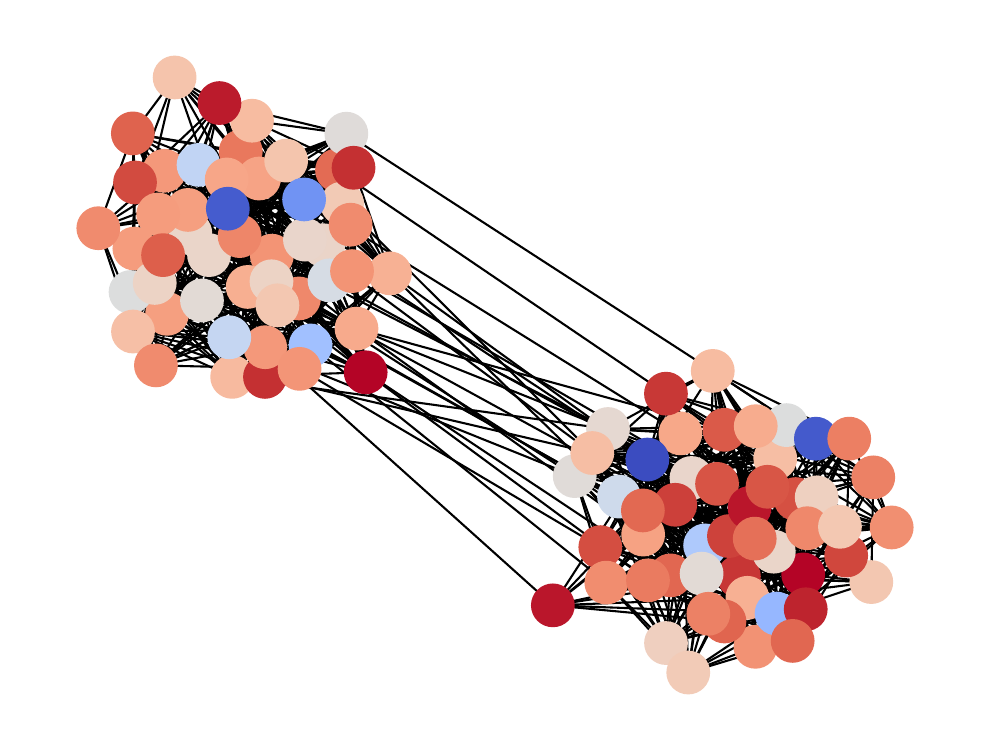}
\caption{$\alpha_\Lambda=15, \beta_\Lambda=2$}
\end{subfigure}
\begin{subfigure}[b]{0.13\linewidth}
    \includegraphics[width=\linewidth]{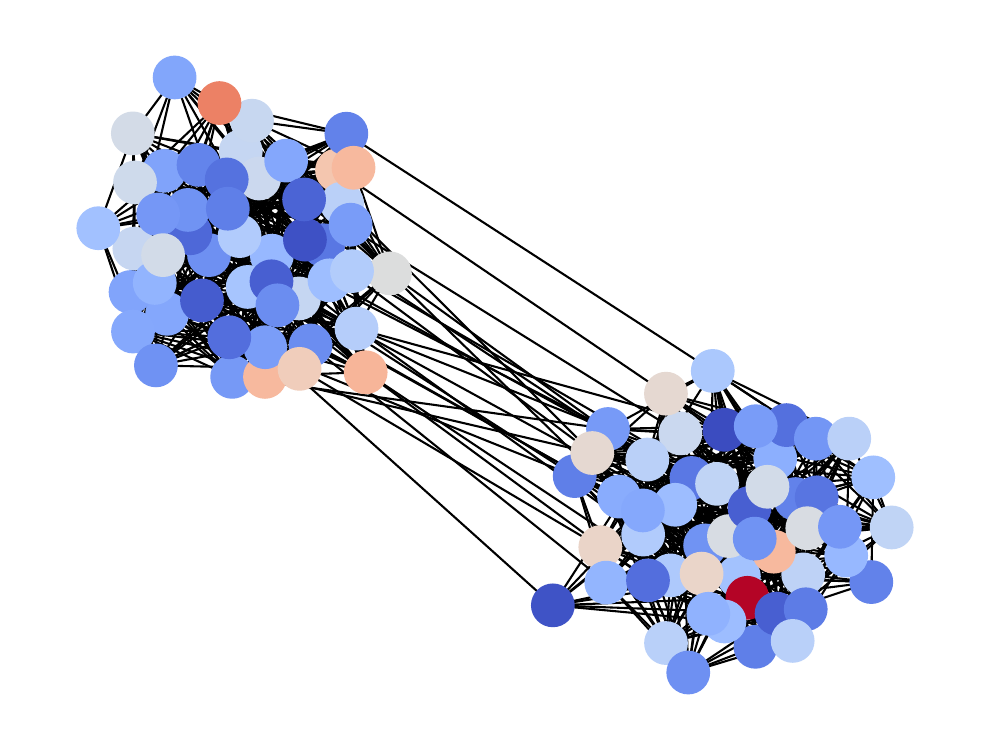}
    \caption{$\alpha_\Lambda=2, \beta_\Lambda=5$}
\end{subfigure}
\begin{subfigure}[b]{0.13\linewidth}
\includegraphics[width=\linewidth]{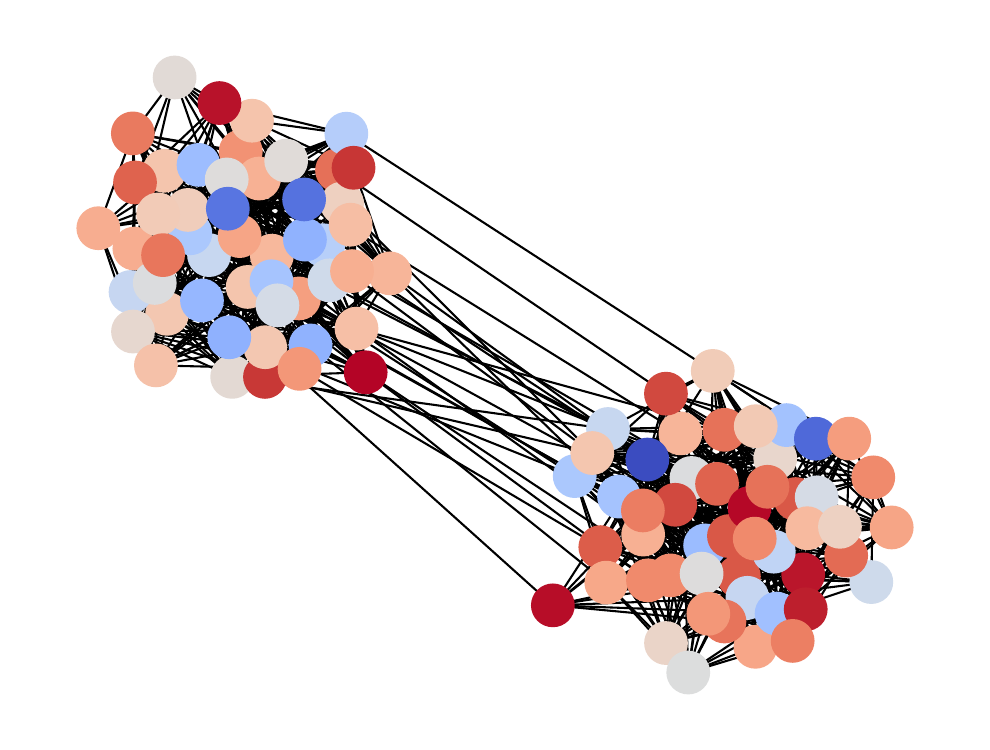}
\caption{$\alpha_\Lambda=5, \beta_\Lambda=2$}
\end{subfigure}
\begin{subfigure}[b]{0.13\linewidth}
\includegraphics[width=\linewidth]{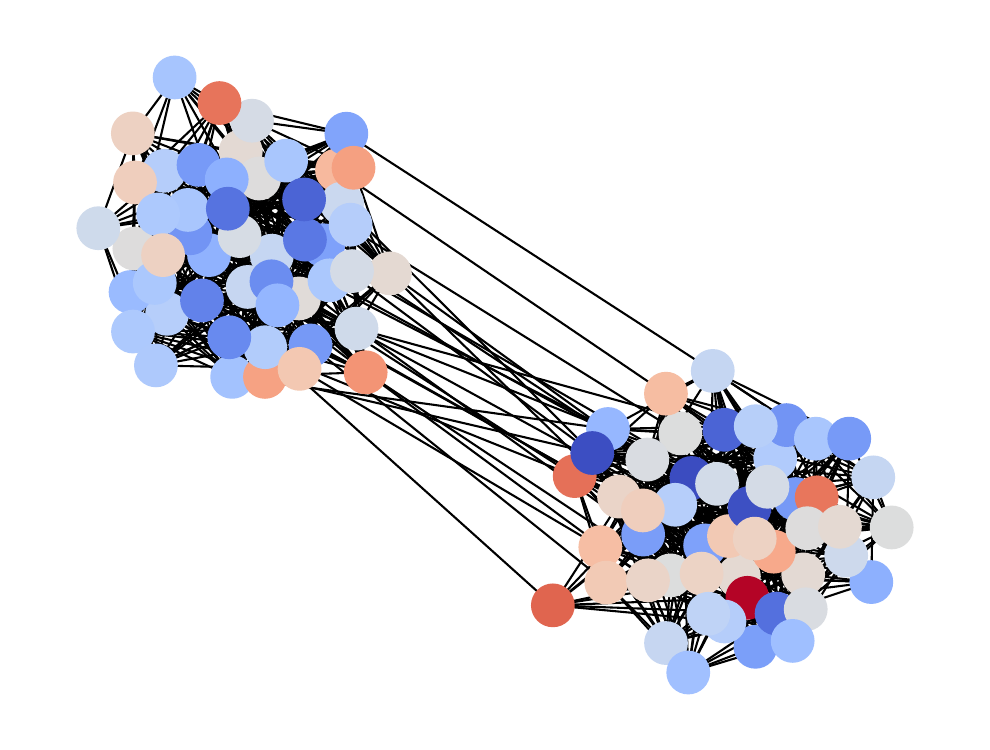}
\caption{$\alpha_\Lambda=15, \beta_\Lambda=15$}
\end{subfigure}
\begin{subfigure}[b]{0.13\linewidth}
\includegraphics[width=\linewidth]{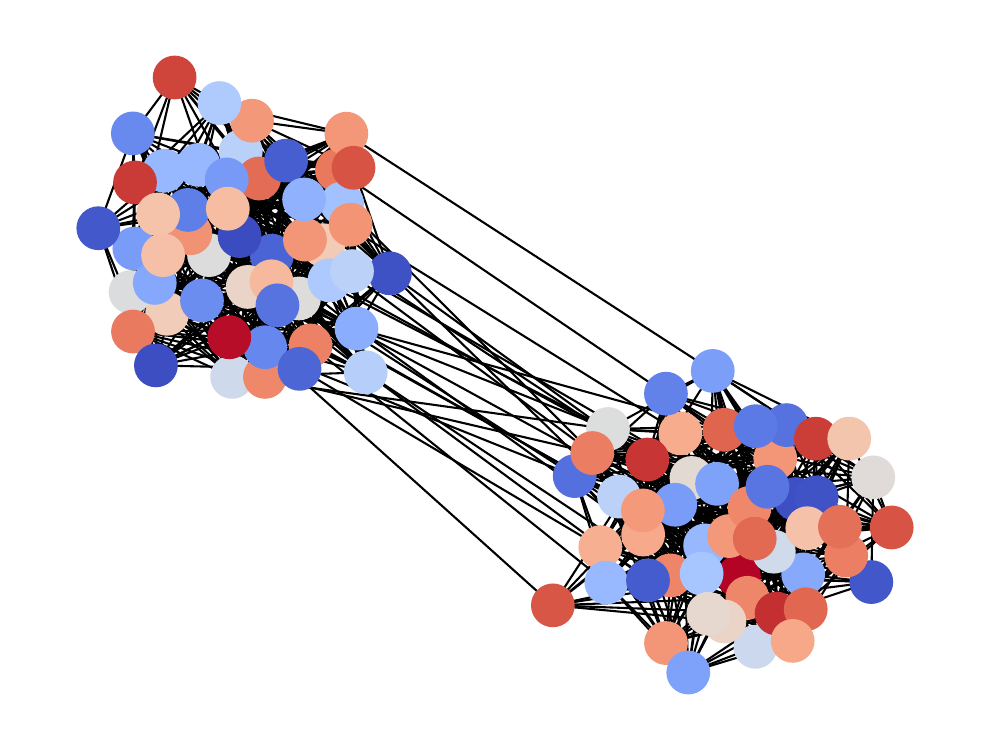}
\caption{$\alpha_\Lambda=1, \beta_\Lambda=1$}
\end{subfigure}
    \caption{Susceptibility distribution on the synthetic network by varying the Beta parameters, from highly skewed ((a)-(b)) to slightly skewed ((c)-(d)) to uniform ((e)-(f)).}
    \label{fig:synthetic-networks-lambda}
\end{figure*}

\section{Evaluation}
\label{sec:setup}
We evaluate the capabilities of the framework in disrupting a network. First, we empirically validate the findings and the approach discussed in Section~\ref{sec:model}. Next, we show that the content generated by Algorithm~\ref{alg:procedure} effectively maximizes disruption. 

\subsection{Experimental setup}
In this section, we discuss the data used in the experimental evaluation of the framework and the implementation details.

\paragraph{Synthetic Networks}
{We begin our evaluation using synthetic graphs. These graphs are generated via a stochastic block model~\cite{HOLLAND1983109}, partitioning the nodes in two blocks of arbitrary size, and placing edges between pairs of nodes with a probability that depends on the blocks. In our experiments, we set the size of each community equal to $50$, the probability of intracommunity edges equal to $0.4$, and the probability of intercommunity edges equal to $0.01$. 
Following previous approaches~\cite{cinus2022effect, coppolillo2025engagementdrivencontentgenerationlarge}, we generate the innate opinion vector \innateopinions via a Beta distribution, i.e., $\innateopinion_i \sim \textrm{Beta}(\alpha, \beta)$. Specifically, we vary the $\alpha, \beta$ parameters in the range 
{$[1, 5, 10, 15]$}
over each community, denoting them as ($\alpha_1, \beta_1$) and ($\alpha_2, \beta_2$), respectively. We fix $\alpha_2 = \beta_1 = 15$ and vary $\alpha_1$ and $\beta_2$ to control the homophily of the network, from high to low, as reported in Figure~\ref{fig:synthetic-networks-opinions}.

Similarly, we generate the user susceptibility values by using a Beta distribution, i.e., $\Lambda \sim$ Beta($\alpha_\Lambda, \beta_\Lambda$), spanning $\alpha_\Lambda, \beta_\Lambda$ in the range $[1, 2, 5, 15]$ to vary the susceptibility from skewed (on either $0$ or $1$) to uniformly distributed. See Figure~\ref{fig:synthetic-networks-lambda} for reference.

Regarding the influence matrix $W$, we assume the following. Given an edge $(i, j)$, if the sign of $\innateopinion_i$ differs from the sign $\innateopinion_j$ and $|\innateopinion_i - \innateopinion_j| \geq \varepsilon$, then $w_{ij} = -1$; otherwise, $w_{ij} = 1$. We further assume $w_{ij} = 0$ if no connection exists between nodes $i$ and $j$. Concerning $\varepsilon$, higher values imply a reduction of negative weights in the network, and a consequent increase of positive influence. In the Results section, we provide an exhaustive experimentation on how different values of $\innateopinions, \Lambda$ and $\varepsilon$ affect the results in terms of induced disruption.

\paragraph{Real-Wold Networks}
In addition to synthetic graphs, we evaluate our approach on two real-world datasets from the $\mathbb{X}$ social network.

\begin{itemize}[leftmargin=*]
    \item \textbf{Brexit}: centered around the 2016 UK referendum on European Union membership~\cite{ZHU2020102031}. We use the version presented in~\cite{minici2022cascade}, comprising 7,589 users, 532,459 links, and 19,963 tweets annotated with binary stance labels. Each user is assigned an opinion value $s_u \in [-1,1]$, representing the average stance of the tweets they retweeted ($-1$ for "Remain", $1$ for "Leave").

\item \textbf{Italian Referendum}: focused on the 2016 Italian constitutional referendum~\cite{lai2018stance}, processed analogously to the Brexit data~\cite{minici2022cascade}. It contains 2,894 users, 161,888 edges, and 41,001 tweets. As in Brexit, $s_u \in [-1,1]$ ($-1$ for “No”, $1$ for “Yes”).
\end{itemize}



As illustrated in Figure~\ref{fig:real_networks} (Appendix), both datasets display strongly homophilic and polarized communities.
The influence matrix $W$ is generated as in the synthetic case, while the susceptibility matrix $\Lambda$ is assumed to be proportional to the nodes degree. 
We argue that this specific equilibrium holds because the ``stubbornness'' of each agent is implicitly tied to their degree: an agent with many neighbors (high \(d_{i}\)) is effectively more susceptible to others influence, while an agent with few neighbors is relatively more influenced by their own innate opinion. 
We acknowledge that more sophisticated approaches for estimating $W$ and $\Lambda$ from the underlying network structure and user activity data could be employed~\cite{minici2022cascade,10.1145/3459637.3482444,10.1145/2623330.2623733}. We leave the investigation of such refined formulations for future work.



\paragraph{Settings}
We instantiate the scoring function $\mathcal{S}$ with a RoBERTa-base model trained on $\sim$124M tweets,
and fine-tuned for sentiment analysis and stance detection with the TweetEval benchmark~\cite{loureiro-etal-2022-timelms, barbieri-etal-2020-tweeteval}. The  average F-score reported on the stance detection task is equal to $72.9$, thus ensuring reliability. 

As aforesaid, the prompt used to generate queries is of the form: “Generate a post about \texttt{[TOPIC]}”, where \texttt{[TOPIC]} is chosen based on the network (“cats” for synthetic networks, and “Brexit” or “the 2016 Italian Referendum” for real-world cases).
For the experiments, we employ the 2B parameter variant of Gemma~\cite{gemma}, a compact large language model based on Google’s Gemini architecture.\footnote{\url{https://deepmind.google/technologies/gemini/}} Despite its relatively small size, Gemma-2B demonstrates competitive performance on tasks involving reasoning, comprehension, and safety. To align the model-generated statements with the desired stance, we fine-tune Gemma using the \texttt{PPOTrainer} class from the \texttt{trl} library,\footnote{\url{https://huggingface.co/docs/trl/main/en/ppo_trainer\#trl.PPOTrainer}} which enables reinforcement learning with custom reward functions. We set the maximum training steps to 10,000 and batch size equal to 8, with a KL-threshold $\tau = 50$.
For fine-tuning, we used 1 GPU NVIDIA A100-SXM4-80GB. 

Our approach can be feasibly applied to large graphs, since $\expressedopinions^*$ can be computed efficiently even when $W$ is large~\cite{10591448}.
%
We further emphasize that our framework is model-agnostic and can accommodate \textbf{any} LLM, provided that it is amenable to fine-tuning. 

\subsection{Results}
\label{sec:results}
\paragraph{Disruption on Synthetic Graphs} First, we exploit the synthetic networks to validate the theoretical findings provided in Section~\ref{sec:model} by comparing: the disruption obtained at the initial opinion state 
($I_{\graph,\innateopinions}$), the disruption at equilibrium computed with the basic FJ model ($I_{\graph,\expressedopinions_\textrm{Basic}^*}$), the disruption at equilibrium computed with the extended FJ model ($I_{\graph,\expressedopinions_\textrm{Extended}^*}$), and finally the disruption at equilibrium computed with the extended FJ model and the opinion manipulated according to Equation~\ref{eq:problem-formulation} ($I_{\graph,\expressedopinions_\textrm{Manipulated}^*}$). Table~\ref{tab:synth_results_opinions_param} reports the results on the synthetic networks by varying the Beta parameters to gradually decrease the graph homophily (see Figure~\ref{fig:synthetic-networks-opinions} as reference). Here, the parameters for generating $\Lambda$ are fixed to $\alpha_\Lambda = \beta_\Lambda = 1$ and $\varepsilon = 0$. {For robustness, we generate the graphs with $5$ different seeds, and average the corresponding scores.}

        \begin{table}[!ht]
        \centering
        \caption{Results on the synthetic network by varying the opinion distribution (see Figure~\ref{fig:synthetic-networks-opinions}), computed in terms of: disruption at initial opinion state ($I_{\graph,\innateopinions}$), disruption at equilibrium via the basic FJ model ($I_{\graph,\expressedopinions_\textrm{Basic}^*}$), disruption at equilibrium via the extended FJ model ($I_{\graph,\expressedopinions_\textrm{Extended}^*}$), and disruption at equilibrium via the extended FJ model with the manipulated opinion ($I_{\graph,\expressedopinions_\textrm{Manipulated}^*}$ ). {Scores are averages across $5$ runs.}}
        \label{tab:synth_results_opinions_param}
        \resizebox{0.8\columnwidth}{!}{
        \begin{tabular}{ccccc}
        \toprule
            Parameters & $I_{\graph, \innateopinions}$ & $I_{\graph, \expressedopinions^*_{\textrm{Basic}}}$ & $I_{\graph, \expressedopinions^*_{\textrm{Extended}}}$ & 
            $I_{\graph, \expressedopinions^*_{\textrm{Manipulated}}}$\\
            \midrule
        $\alpha_1 = \beta_2 = 1$ & 328.45 & 176.12 & 2477.1 & 5858.8\\ 
$\alpha_1 = \beta_2 = 5$ & 576.39 & 464.81 & 1995.91 & 4187.58\\ 
$\alpha_1 = \beta_2 = 10$ & 332.19 & 249.02 & 472.99 & 1385.32\\ 
$\alpha_1 = \beta_2 = 15$ & 309.3 & 241.56 & 158.26 & 545.03\\ 

            \bottomrule
        \end{tabular}
        }
        \end{table}

We notice that the value of $I_{\graph, \expressedopinions_\textrm{Basic}^*}$ is always significantly lower than $I_{\graph, \innateopinions}$, independently of the opinion configuration. This confirms our theoretical results, showing that the standard version of the FJ model prevents disrupting the underlying network. Contrarily, adopting the extended version of the FJ model enables a significant increase in the induced disruption, which is further maximized by applying opinion manipulation. Notably, we see that such disruption correlates with the innate opinion distribution of the nodes: 
the higher the homophily, the higher the disruption on the graph. 

Next, we analyze how the user susceptibility affects the induced disruption. Table~\ref{tab:synth_results_lambda} reports the scores on the synthetic networks, from skewed to uniformly distributed susceptibility among users. The Beta parameters used to generate the opinions are here equal to $\alpha_1 = \beta_2 = 1$ and $\varepsilon = 0$. {Again, we generated the underlying graph with $5$ different seeds and average the scores for ensuring robustness.} 
{The results suggest that the induced disruption correlates with the user susceptibility, progressively increasing as the underlying distribution becomes uniform, mixing together low- and high-susceptible users. This is a reasonable outcome, since opinion disruption would be unlikely to occur if most users are highly susceptible ($\alpha_\Lambda \in [5, 15]$, $\beta_\Lambda = 2$) or conversely highly stubborn ($\alpha_\Lambda = 2$, $\beta_\Lambda \in [5, 15]$). A mixed distribution is therefore necessary to disrupt the opinions of the underlying graph.}

            \begin{table}[!ht]
            \centering
\caption{Results on the synthetic network by varying the Beta parameters in generating $\Lambda$ (see Figure~\ref{fig:synthetic-networks-lambda}), computed in terms of disruption at equilibrium via the extended FJ model ($I_{\graph,\expressedopinions_\textrm{Extended}^*}$), and via the extended FJ model with the manipulated opinion ($I_{\graph,\expressedopinions_\textrm{Manipulated}^*}$). {Scores are averages across $5$ runs.}}
            \label{tab:synth_results_lambda}
            \resizebox{0.6\columnwidth}{!}{
            \begin{tabular}{ccc}
            \toprule
                Parameters & $I_{\graph, \expressedopinions^*_{\textrm{Extended}}}$ & 
                $I_{\graph, \expressedopinions^*_{\textrm{Manipulated}}}$\\
                \midrule
$\alpha_\Lambda =15, \beta_\Lambda=2$ & 70.19 & 287.71\\ 
$\alpha_\Lambda =5, \beta_\Lambda=2$ & 246.62 & 1070.07\\ 
$\alpha_\Lambda =2, \beta_\Lambda=15$ & 278.8 & 1118.37\\ 
$\alpha_\Lambda =2, \beta_\Lambda=5$ & 222.95 & 1164.9\\ 
$\alpha_\Lambda =15, \beta_\Lambda=15$ & 286.03 & 1206.4\\ 
            $\alpha_\Lambda =1, \beta_\Lambda=1$ & 2477.1 & 5858.8\\ 


                \bottomrule
            \end{tabular}
            }
            \end{table}

Finally, we investigate the impact of $\varepsilon$ in generating the influence weights $W$, as depicted in Table~\ref{tab:synth_results_epsilon}. Here, we use an underlying network exhibiting high homophily and uniformly distributed susceptibility. {As before, we average the scores across $5$ different generation runs.} 
{Interestingly, we notice that increasing $\varepsilon$ beyond a certain threshold leads to higher scores in terms of disruption, suggesting that the underlying graph is maximally disrupted when fewer negative influences are present among nodes.}

             \begin{table}[!ht]
             \centering
\caption{Results on the synthetic network by varying the $\varepsilon$ parameter in generating $W$, computed in terms of disruption at equilibrium via the extended FJ model ($I_{\graph,\expressedopinions_\textrm{Extended}^*}$), and via the extended FJ model with the manipulated opinion ($I_{\graph,\expressedopinions_\textrm{Manipulated}^*}$). {Scores are averages across $5$ runs.}}
             \label{tab:synth_results_epsilon}
             \resizebox{0.42\columnwidth}{!}{
             \begin{tabular}{ccc}
             \toprule
                 $\varepsilon$ & $I_{\graph, \expressedopinions^*_{\textrm{Extended}}}$ & 
                 $I_{\graph, \expressedopinions^*_{\textrm{Manipulated}}}$\\
                 \midrule
             0 & 2477.1 & 5858.8\\ 
0.5 & 2477.1 & 5858.8\\ 
1 & 2477.1 & 5858.8\\ 
1.5 & 2487.39 & 5869.29\\ 
2 & 2777.74 & 6368.77\\ 

                 \bottomrule
             \end{tabular}
             }
             \end{table}

\begin{table}[!ht]
    \centering
    
    \caption{Absolute value scores on the real networks computed in terms of: disruption at initial opinion state ($I_{\graph,\innateopinions}$), disruption at equilibrium via the basic FJ model ($I_{\graph,\expressedopinions_\textrm{Basic}^*}$), via the extended FJ model ($I_{\graph,\expressedopinions_\textrm{Extended}^*}$), and via the extended FJ model with the manipulated opinion ($I_{\graph,\expressedopinions_\textrm{Manipulated}^*}$).}
    \label{tab:real_results}\resizebox{0.8\columnwidth}{!}{
    \begin{tabular}{ccccc}
    \toprule
         Dataset & $I_{\graph, \innateopinions}$ & $I_{\graph, \expressedopinions^*_{\textrm{Basic}}}$ & $I_{\graph, \expressedopinions^*_{\textrm{Extended}}}$ & 
            $I_{\graph, \expressedopinions^*_{\textrm{Manipulated}}}$ \\
        \midrule
        Brexit & 4.083e+09 & 4.082e+09 & 6.810e+12 & 6.828e+12 \\
        Referendum & 9.229e+09 & 9.229e+09 & 3.765e+10 & 3.771e+10 \\
        \bottomrule
    \end{tabular}
    }
\end{table}
\paragraph{Disruption on Real Networks} Now, we move our investigation toward real-world social graphs. First, we compare the disruption obtained at the inner state and at equilibrium with the basic and extended FJ model, further computing the score by using the manipulated opinion vector according to Equation~\ref{eq:problem-formulation}. Table~\ref{tab:real_results} provides the scores in absolute terms on the Brexit and Italian Referendum datasets, by setting $\varepsilon = 0$. The results are consistent with the ones obtained on synthetic networks, further validating our findings on real-world networks. Next, we investigate the impact of $\varepsilon$ on the induced disruption. For the sake of readability, in Table~\ref{tab:real_results_epsilon}, we provide the scores normalized via min-max scaling.

                     \begin{table}[!ht]
                     \centering
                     \caption{Results on the real-world networks by varying the $\varepsilon$ parameter in generating $W$, computed in terms of disruption at equilibrium via the extended FJ model ($I_{\graph,\expressedopinions_\textrm{Extended}^*}$), and via the extended FJ model with the manipulated opinion ($I_{\graph,\expressedopinions_\textrm{Manipulated}^*}$). Scores are normalized via min-max scaling.}
                     \label{tab:real_results_epsilon}
                     \resizebox{0.75\columnwidth}{!}{
                     \begin{tabular}{ccccc}
                     \toprule
                         $\varepsilon$ & \multicolumn{2}{c}{Brexit} & \multicolumn{2}{c}{Referendum} \\
                         \cmidrule(lr){2-3} 
                         \cmidrule(lr){4-5} 
                         &
                         $I_{\graph, \expressedopinions^*_{\textrm{Extended}}}$ & 
                         $I_{\graph, \expressedopinions^*_{\textrm{Manipulated}}}$ & 
                         $I_{\graph, \expressedopinions^*_{\textrm{Extended}}}$ & 
                         $I_{\graph, \expressedopinions^*_{\textrm{Manipulated}}}$ \\
                         \midrule
                     0 & 0.997 & 1.0 & 0.856 & 0.858\\
                     0.5 & 0.976 & 0.979 & 0.856 & 0.858\\
                     1 & 0.7 & 0.703 & 0.855 & 0.857\\
                     1.5 & 0.534 & 0.537 & 0.908 & 0.91\\
                     2 & 0.265 & 0.269 & 0.998 & 1.0\\
                 \bottomrule
             \end{tabular}
             }
             \end{table}
Notably, we devise an opposite trend comparing the two networks: while the induced disruption is positively correlated to $\varepsilon$ on the Italian Referendum dataset, it decreases when $\varepsilon$ increases on Brexit. 
{We speculate that this can be due to the properties of the Brexit graph, which displays less homophily than the Italian Referendum, thus resulting in more negative influence among nodes.}

Further, we analyze how the location of the node $u$ in the graph affects the induced disruption after manipulating $\innateopinion_u$. 
Specifically, we perform a correlation analysis between the centrality of the node and the produced disruption. As centrality metrics, we considered node degree, betweenness centrality~\cite{BRANDES2008136}, and eigenvector centrality~\cite{bonacich1972technique}.  
{We found that the eigenvector centrality is the only measure displaying a moderate correlation with the induced disruption on the graph: positive on Brexit, and negative on the Italian Referendum network. Conversely, the degree and the betweenness centrality do not show any significant correlation. We hypothesized that this depends on the metric property, which estimates a node importance through its connections to \textit{influential} nodes. Therefore, being more or less connected to influential users can affect the outcome in terms of disruption (see Figure~\ref{fig:disruption-centrality} and Table~\ref{tab:correlation}).
}


\begin{table*}[!ht]
    \centering
    
    \caption{Texts generated at convergence along with the corresponding stance, reward, and target $\innateopinion'_u$.}
    \label{tab:results}
    \resizebox{0.8\linewidth}{!}{
    \begin{tabular}{c c p{9cm} c c}
    \toprule
        \textbf{Network}  & \textbf{Target} $\innateopinion'_u$ & \textbf{Generated Text} &  \textbf{Stance} & \textbf{Reward} \\
        \midrule
        Synthetic-positive & 1 & \textit{A fluffy ball of purring energy with soft, warm fur and a mischievous grin, seemed to find its way into my lap and never left, leaving me with only the sweetest memories to cherish.}
 & 0.963 & 0.872 \\
         \cmidrule(lr){1-5}

        Synthetic-negative & -1 & \textit{Cats are vermin, a nuisance that has no place in a civilized society.} & -0.954 & 0.977\\
                \cmidrule(lr){1-5}

        Brexit & 1 & \textit{Despite the tumultuous political landscape, the iconic red phone became a symbol of unity and resilience, as people from all walks of life rallied together to celebrate the country's post-Brexit legacy.} & 0.967 & 0.988 \\
        \cmidrule(lr){1-5}
        Referendum & -1 & \textit{The horrific 2016 Italian referendum illustrated the devastating impact political polarization can have on society.}
        & -0.951 & 0.974 \\
    \bottomrule
    \end{tabular}
    }
\end{table*}
\begin{figure}[!ht]
    \centering
    \includegraphics[width=0.73\linewidth]{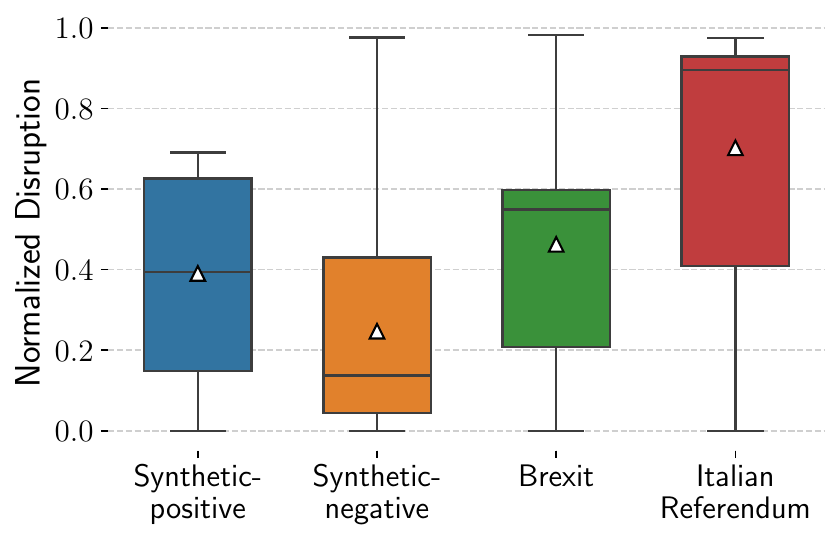}
    \caption{Distribution of the (normalized) disruption scores induced by the generated texts. Triangles indicate means.}
    \label{fig:disruption-generated}
\end{figure}
\paragraph{Generate Disruptive Content} Finally, we discuss the effectiveness of our LLM fine-tuning framework in generating disruptive text. 
To this end, for each considered network, we select the injecting node $u$ whose manipulated opinion $\innateopinion'_u$ maximally disrupts the graph according to Equation~\ref{eq:problem-formulation}. Regarding the synthetic networks, we consider two configurations where $\innateopinion'_u \in \{-1, 1\}$, in order to evaluate both a negative and a positive target. We refer to these networks as ``Synthetic-negative'' and ``Synthetic-positive'', respectively.

We preliminarily analyze the disruption distribution induced by the texts generated during training (assuming equal $u$).
Figure~\ref{fig:disruption-generated} depicts the results. The Y-axis refer to the disrupted scores normalized via min-max normalization, where $1$ denotes the theoretical disruption upper-bound. 
We notice that, in all networks (except for the Italian Referendum), most of the generated content induces a low score, especially in the case of the synthetic network where the target opinion is negative. We suppose that the higher scores induced over the Italian Referendum graph are due to ({i}) the fact that the most users have negative stance towards the referendum and ({ii}) the presence of a consistent cluster of users with mixed opinions, which can better conveys disruption (see Figure~\ref{fig:referendum-viz}). Nevertheless, all four distributions show that it is not trivial to craft a message that maximizes disruption over the networks. 
\begin{figure}[!ht]
    \centering
\includegraphics[width=0.8\linewidth]{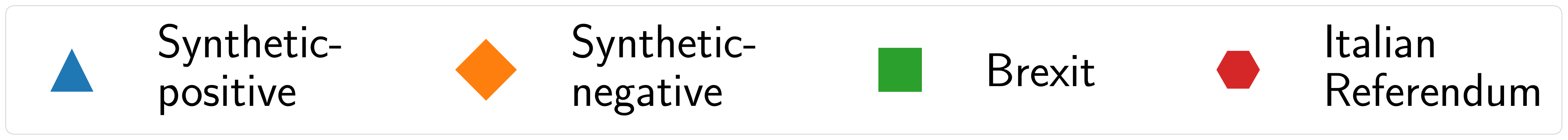}
\includegraphics[width=\linewidth]{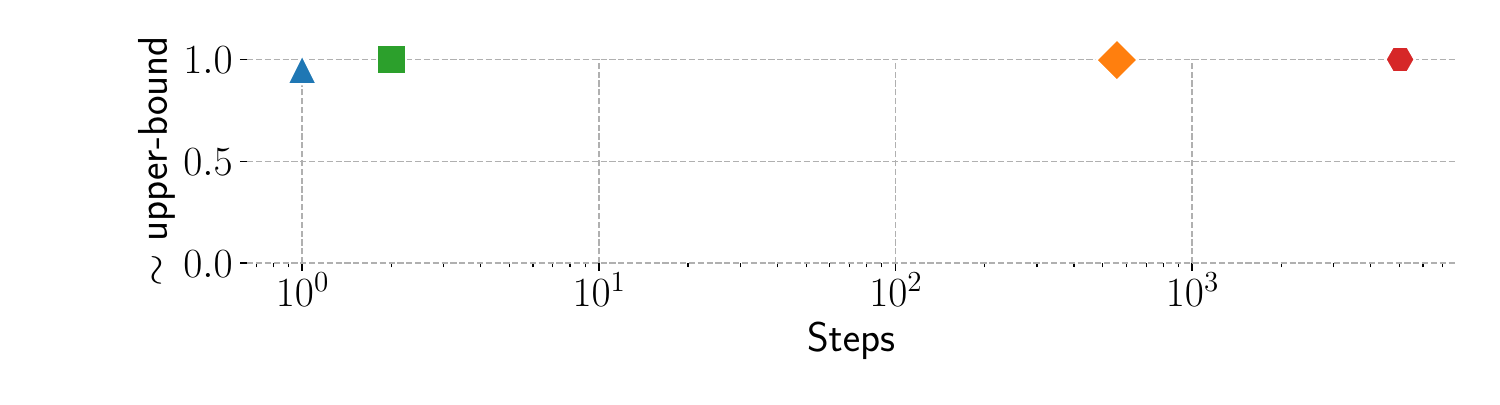}
    \caption{Correlation between the number of training steps needed for convergence (X-axis) and the proximity of the obtained disruption to the network upper-bound (Y-axis).}
    \label{fig:disruption-convergence}
\end{figure}
\\
Further, Table~\ref{tab:results} provides the texts generated at convergence, with the corresponding stance and model reward for each network. We notice that the generated content  maintains a realistic human-like fashion; further, the associated stance approaches (by reward design) the target disruptive opinion $\innateopinion'_\node$. 
Additionally, we depict in Figure~\ref{fig:disruption-convergence} the correlation between the disruption score induced by the final generated text and the number of steps needed to reach convergence. Specifically, the X-axis reports the number of steps when the stopping condition occurred, i.e., $x \approx \innateopinion'_u$, (see Algorithm~\ref{alg:procedure}), while the Y-axis represents the proximity of the disruption induced by the generated content to the network \textit{upper-bound}, i.e., the highest disruption achievable.
Two considerations can be made. First, in all social graphs, the disruption induced by generated texts almost perfectly approaches the upper-bound.
Second, while the procedure takes a few steps to reach convergence on the positive-oriented networks (Synthetic-positive and Brexit), it requires a significantly higher number of steps on the negative ones. This aligns with previous studies assessing how current LLMs are prone to generate positive-sentiment texts while being more difficult to fine-tune for negative content~\cite{coppolillo2025engagementdrivencontentgenerationlarge}.
We finally provide convergence time: the Referendum dataset requires the highest amount of time, which is $\sim$13h, followed by Synthetic-negative ($\sim$1.5h), Brexit ($\sim$20') and Synthetic-positive ($\sim$10').
\section{Conclusions and Future Work}
\label{sec:conclusions}
In this work, we examined how social disruption can be strategically amplified in online networks. Extending the FJ model with node susceptibility and directional influence, we showed that targeted opinion perturbations can effectively amplify social dissensus.
To this end, we fine-tuned a Large Language Model via RL, generating maximally disruptive content. Experiments on synthetic and real networks show that LLM-induced disruption closely tracks theoretical bounds. Nonetheless, our approach can be extended in several directions. First, reward modeling can be refined by integrating other factors beyond stance (e.g., writing style, social dimensions). Secondly, user representation could be enriched by considering multiple texts, instead of a single post. Finally, the function $\mathcal{S}$ that we operationalized through a stance classifier can be, in principle, instantiated via any reliable source (e.g., human annotators, an LLM-as-a-judge). Stance estimation is indeed orthogonal to the proposed framework, and can be operazionalized in different ways. 

\section*{Acknowledgments}
This project has been partially funded by Moneying-Plus project (CUP J29I24001790005) selected within the framework of the PR FESR – FSE Calabria 2021/2027 and implemented with the support of the Italian State and the Calabria Region – Action 1.1.1: Support for research, development, and innovation projects, including those carried out in collaboration with research organizations, within the priority areas and trajectories of the S3 Strategy.
\clearpage

\appendix

\section{Proofs}\label{sec:proof}
\paragraph{FJ Model.}
\label{eq:appendix-fj-model}
We recall the basic FJ dynamics in matrix form:
\begin{equation}
    \begin{split}
    \expressedopinions(0) & = \innateopinions\\
    \expressedopinions(t+1) & = (I + D)^{-1} \left(\innateopinions + A  \expressedopinions(t) \right)
    \end{split}
\end{equation}

By Gershgorin Circle Theorem~\cite{horn13}, $\rho\left(M\right) < 1$ where 
 $M=(I + D)^{-1}A$. As a consequence, the above equation converges. 
Further, at convergence, $\expressedopinions^*$ satisfies the fixed-point equation 
$
\expressedopinions  = (I + D)^{-1} \left(\innateopinions + A  \expressedopinions \right)
$, 
which resolves into
$
(I + D - A)\expressedopinions^* = \innateopinions.
$
Since $L=D-A$, we can rewrite the system as $(I+L)\expressedopinions^* = \innateopinions$. It can be shown that the matrix $I + L$ is invertible. As a result, we obtain the following.


\begin{theorem}
    The $FJ$ model converges to  $\expressedopinions^* = (I+L)^{-1}\innateopinions$. 
\end{theorem}
\keyLemma*
\begin{proof}
Let $\gamma=-\frac{1}{n}$, $\tilde{L}=I+L$, and $
H \;=\; \tilde{L} + \gamma\mathbf{1}\mathbf{1}^\top
\;=\; I + L - \frac{1}{n}\mathbf{1}\mathbf{1}^\top.$
Since $\expressedopinions^*=\tilde{L}^{-1}\innateopinions$ (Appendix~\ref{sec:proof}), we can write
\[
\disruption_{\graph,\expressedopinions^*}=\innateopinions^\top\tilde{L}^{-1}H\tilde{L}^{-1}\innateopinions,
\qquad
\disruption_{\graph,\innateopinions}=\innateopinions^\top H\innateopinions.
\]
\[
\implies
\disruption_{\graph,\expressedopinions^*}-\disruption_{\graph,\innateopinions}
= \innateopinions^\top Y\,\innateopinions,
\qquad
Y := \tilde{L}^{-1}H\tilde{L}^{-1}-H.
\]
It is thus sufficient to show that $Y\preceq 0$.

\textbf{Step 1 (rewrite $Y$).} Expanding the definition of $H$ gives
\begin{equation}
\begin{aligned}
Y
&= \tilde{L}^{-1}(\tilde{L}+\gamma\mathbf{1}\mathbf{1}^\top)\tilde{L}^{-1}-(\tilde{L}+\gamma\mathbf{1}\mathbf{1}^\top)\\
&= \tilde{L}^{-1}-\tilde{L}+\gamma\tilde{L}^{-1}\mathbf{1}\mathbf{1}^\top\tilde{L}^{-1}-\gamma\mathbf{1}\mathbf{1}^\top.
\end{aligned}
\label{eq:y}
\end{equation}

\textbf{Step 2 (eigenpair for $\mathbf{1}$).} Since $L\mathbf{1}=\mathbf{0}$, we have
\[
\tilde{L}\mathbf{1}=(I+L)\mathbf{1}=\mathbf{1}
\quad\Rightarrow\quad
\tilde{L}^{-1}\mathbf{1}=\mathbf{1}.
\]
Moreover, $\mathbf{1}\mathbf{1}^\top\mathbf{1}=n\mathbf{1}$. Plugging into~\eqref{eq:y} yields
\[
Y\mathbf{1} = (\mathbf{1}-\mathbf{1}) + \gamma(n\mathbf{1}-n\mathbf{1}) = \mathbf{0},
\]
so $\mathbf{1}$ is an eigenvector of $Y$ with eigenvalue $0$.

\textbf{Step 3 (orthogonal eigenvectors).} Let $\{\mathbf{v}_1,\dots,\mathbf{v}_n\}$ be an orthonormal eigenbasis of $\tilde{L}$ with
$\mathbf{v}_1=\mathbf{1}/\sqrt{n}$ and eigenvalues $1=\sigma_1\leq\sigma_2\leq\cdots\leq\sigma_n$. (the latter holds since $\sigma_i - 1$ is an eigenvalue of $L$ and the latter is positive semidefinite, as shown in lemma~\ref{lemma:centeringmat}). 
For $j\geq 2$ we have $\mathbf{1}^\top\mathbf{v}_j=0$, hence
$\mathbf{1}\mathbf{1}^\top\mathbf{v}_j=\mathbf{0}$ and also
$\mathbf{1}\mathbf{1}^\top\tilde{L}^{-1}\mathbf{v}_j=\mathbf{0}$.
Thus, using~\eqref{eq:y}, for all $j\geq 2$,
\[
Y\mathbf{v}_j = \tilde{L}^{-1}\mathbf{v}_j - \tilde{L}\mathbf{v}_j
= \Bigl(\frac{1}{\sigma_j}-\sigma_j\Bigr)\mathbf{v}_j.
\]
Since $\tilde{L}=I+L$ and $L\succeq 0$, we have $\sigma_j\geq 1$ for $j\geq 2$, so
$\frac{1}{\sigma_j}-\sigma_j\leq 0$.
Therefore all eigenvalues of $Y$ are non-positive, i.e., $Y\preceq 0$, concluding the proof.
\end{proof}

\paragraph{Properties of the Extended FJ model.}

The opinion dynamics for the extended (signed, weighted, directional) model can be expressed in matrix form as follows: 
\begin{equation}\label{eq:extendedfj}
\begin{split}
\expressedopinions(0) = &\ \innateopinions\\
\expressedopinions(t+1) = & \ (I - \Lambda) \innateopinions + \Lambda D^{-1} W \expressedopinions(t)    
\end{split}
\end{equation}
Since $W$ is not symmetric and can express negative entries, the expressed opinion does not necessarily converge to a steady state.

In the following, we analyze the conditions for convergence and characterize the resulting opinion vector. 

\begin{lemma}\label{eq:spectra_weighted}
Let $P=\Lambda D^{-1}W$. Then 
$\rho\left(P\right) < 1$ if any of the above conditions holds: 
\begin{enumerate}
    \item\label{cond:strictstubborn} (strict stubbornness) $\lambda_i<1$ for each node $i$
    \item\label{cond:mixedsusc} (mixed susceptibility) the subgraph of nodes $i$ such that  $\lambda_i = 1$ is connected to at least one stubborn node $k$ such that $\lambda_k < 1$. 
\end{enumerate}
\end{lemma}
\begin{proof}
By definition,
\[
p_{i,j} = \frac{\lambda_i w_{i,j}}{d_i},
\]
where $0\leq\lambda_i\leq 1$ and $w_{i,j}\in\{-1,0,1\}$. Hence, for every node $i$,
\[
\sum_j |p_{i,j}| = \frac{\lambda_i}{d_i}\sum_j |w_{i,j}| = \lambda_i \leq 1.
\]
By Gershgorin's theorem~\cite{horn13}, for each eigenvalue $\sigma$ of $P$ there exists an index $i$ such that
\[
|\sigma - p_{i,i}| \leq \sum_{j\neq i} |p_{i,j}|,
\]
\[
\implies
|\sigma| \leq |p_{i,i}| + \sum_{j\neq i} |p_{i,j}| \leq \sum_j |p_{i,j}| = \lambda_i.
\]

\textbf{Case 1 (strict stubbornness).}
If condition~\ref{cond:strictstubborn} holds, then $\lambda_i<1$ for all $i$, hence $|\sigma|<1$ for all eigenvalues and $\rho(P)<1$.

\textbf{Case 2 (mixed susceptibility).}
Assume condition~\ref{cond:mixedsusc} holds. Partition $V=S\cup R$ where
$S=\{i:0\leq\lambda_i<1\}$ (stubborn) and $R=\{i:\lambda_i=1\}$ (susceptible).
Suppose, for contradiction, that $P\genericvec=\sigma\genericvec$ with $|\sigma|\geq 1$.
Let $\tildex=\max_k |x_k|>0$ and choose $i$ such that $|x_i|=\tildex$.
From $\sigma x_i=\sum_j p_{i,j}x_j$ we get
\begin{equation}\label{eq:chaineq}
|\sigma|\,\tildex = |\sigma|\,|x_i| \leq \sum_j |p_{i,j}|\,|x_j| \leq \sum_j |p_{i,j}|\,\tildex = \lambda_i\,\tildex.
\end{equation}
Hence $1\leq |\sigma|\leq \lambda_i$. In particular, $i\notin S$ and thus $i\in R$, so $\lambda_i=1$ and $|\sigma|=1$.




Plugging such values in the chain of inequalities expressed by eq.~\ref{eq:chaineq} forces $|x_j|=\tildex$ for every neighbor $j$ of $i$ with $p_{i,j}\neq 0$.
Iterating the same argument along any directed edge with nonzero weight, we conclude that every node reachable from $i$ must also satisfy $|x_j|=\tildex$.
In particular, $i$ cannot reach any stubborn node (otherwise we would have $|x_j|=\tildex$ at some $j\in S$, contradicting $|\sigma||x_j|\leq \lambda_j\tildex<\tildex$).
This contradicts condition~\ref{cond:mixedsusc}, which assumes every susceptible node can reach a stubborn one.

Therefore $|\sigma|<1$ for all eigenvalues and $\rho(P)<1$.
\end{proof}

The above theorem states that equation~\ref{eq:extendedfj} admits a steady state $\expressedopinions^*$, which is the solution to the equation
$$
\expressedopinions =  \ (I - \Lambda) \innateopinion + \Lambda D^{-1} W \expressedopinions
\implies
(I - \Lambda D^{-1} W) \ \expressedopinions = (I - \Lambda) \innateopinion 
$$

\begin{lemma}\label{eq:inverse_weighted}
    Let $P=\Lambda D^{-1}W$. The matrix $I- P$ is invertible if strict stubbornness or mixed susceptibility conditions (as expressed by lemma~\ref{eq:spectra_weighted}) hold.  
\end{lemma}
\begin{proof}
    By lemma~\ref{eq:spectra_weighted}, we know that $\rho(P) < 1$. Let $\sigma$ be an eigenvalue of $P$, with eigenvector \genericvec. Since
    \begin{equation}
        (I-P)\genericvec = I \genericvec - P \genericvec = \genericvec - \sigma \genericvec = (1-\sigma) \genericvec
    \end{equation}
    we have that $1 - \sigma$ is an eigenvalue of $I-P$. Since $\rho(P) < 1$, it trivially follows that $1 - \sigma \neq 0$. 
\end{proof}

By combining lemmas~\ref{eq:spectra_weighted} and~\ref{eq:inverse_weighted}, we can state the following. 

\begin{theorem}\label{eq:fixpoint}
    If strict stubbornness or mixed susceptibility conditions hold, then the solution to equation~\ref{eq:extendedfj} is given by 
    $$
    \expressedopinions^* = (I - \Lambda D^{-1} W)^{-1} \ (I - \Lambda) \innateopinion 
    $$
\end{theorem}
\paragraph{Disruption injection bounds.}

\begin{lemma}\label{lemma:centeringmat}
The matrix $H= I + L - \frac{1}{n} \mathbf{1}\mathbf{1}^T$ is symmetric and positive semidefinite.
\end{lemma}
\begin{proof}
Let $H=L+S$, where $S = I - \frac{1}{n}\mathbf{1}\mathbf{1}^T$.

\textbf{(1) $L\succeq 0$.} Since 
\[
\genericvec^\top L\genericvec = \sum_{(i,j)\in\edges} (x_i-x_j)^2 \geq 0
\quad\text{for all }\genericvec,
\]
so $L$ is positive semidefinite.

\textbf{(2) $S\succeq 0$.} Note that $S$ is symmetric and idempotent:
\begin{equation}
\begin{aligned}
S^2
&= \Bigl(I-\frac{1}{n}\mathbf{1}\mathbf{1}^T\Bigr)\Bigl(I-\frac{1}{n}\mathbf{1}\mathbf{1}^T\Bigr)
= I - \frac{2}{n}\mathbf{1}\mathbf{1}^T + \frac{1}{n^2}\mathbf{1}\mathbf{1}^T\mathbf{1}\mathbf{1}^T\\
&= I - \frac{2}{n}\mathbf{1}\mathbf{1}^T + \frac{n}{n^2}\mathbf{1}\mathbf{1}^T
= I - \frac{1}{n}\mathbf{1}\mathbf{1}^T
= S.
\end{aligned}
\end{equation}
Since every eigenvalue $\sigma$ of $S$ satisfies $\sigma=\sigma^2$, $\sigma\in\{0,1\}$ and $S\succeq 0$.

\textbf{(3)} Since $L\succeq 0$ and $S\succeq 0$, we conclude $H=L+S\succeq 0$.
\end{proof}
\MaximizationLemma*
\begin{proof}
    Since $H$ is positive semidefinite, $B=M^\top H M$ is also symmetric positive semidefinite. Thus, $\innateopinions'^\top B \innateopinions'$ is convex and $\alpha$ will always coincide with either of its bounds. 
\end{proof}

{Theorem~\ref{theo:maximization} naturally extends to interventions on a \emph{fixed set} of $k$ controlled nodes. Let $S\subseteq \nodes$ be such a set and let $E_S\in\mathbb{R}^{n\times k}$ be the selector matrix whose columns are the basis vectors associated to nodes in $S$. Writing the perturbation as $\innateopinions'=\innateopinions - E_S\boldsymbol{\alpha}$ and enforcing $\innateopinions'\in[-1,1]^n$ yields the box-constrained quadratic program
\[
\max_{\boldsymbol{\alpha}\in[\boldsymbol{\ell},\boldsymbol{u}]} \ (\innateopinions - E_S\boldsymbol{\alpha})^\top B\,(\innateopinions - E_S\boldsymbol{\alpha}),
\]
where $[\boldsymbol{\ell},\boldsymbol{u}]$ encodes the per-node bounds induced by $\innateopinions'\in[-1,1]^n$. 
Since the objective is a \emph{convex} function of $\boldsymbol{\alpha}$, maximizing it over a hyper-rectangle attains an optimum at an \emph{extreme point} of the feasible set; hence, there always exists an optimal solution in which each controlled node is pushed to one of its bounds (each component of $\boldsymbol{\alpha}$ equals either $\ell_i$ or $u_i$). This recovers the same ``extremal'' structure identified by Theorem~\ref{theo:maximization} in the single-node case.}

\section{Experimental Results}

\begin{figure}[!ht]
    \centering
    \begin{subfigure}[b]{0.4\columnwidth}
        \includegraphics[width=\columnwidth]{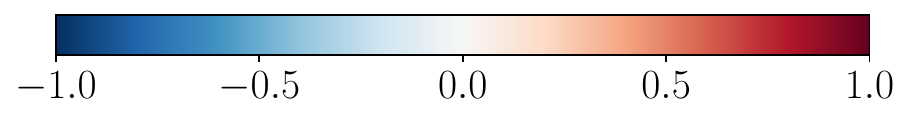}
    \end{subfigure}\\
    \begin{subfigure}[b]{0.2\columnwidth}
    \includegraphics[width=\columnwidth]{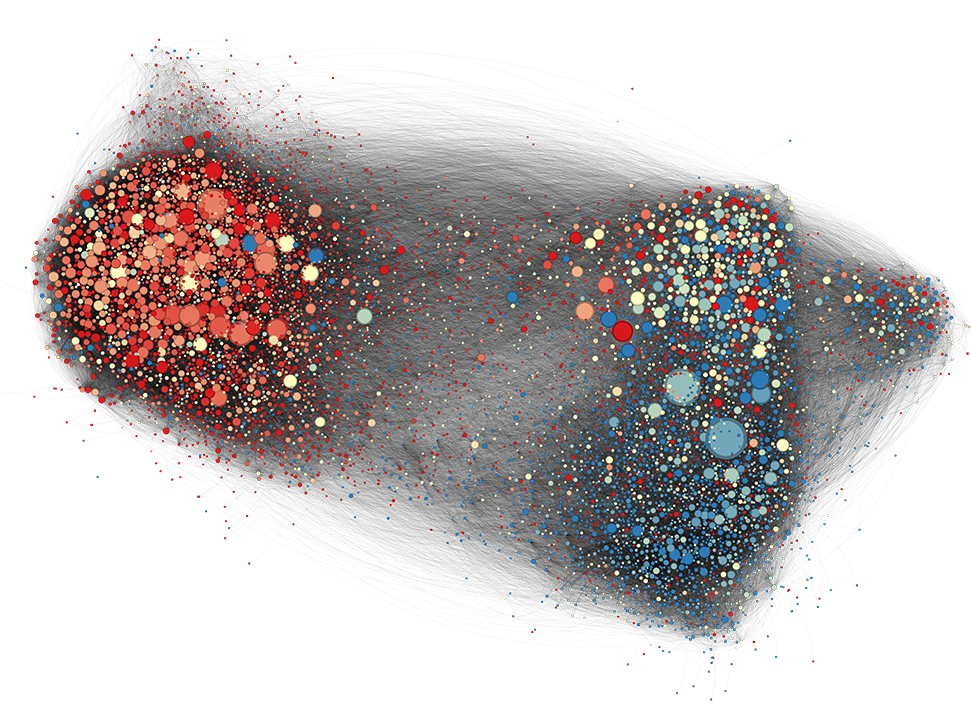}
        \caption{\scriptsize{Brexit}}
        \label{fig:brexit-viz}
    \end{subfigure}
    \quad
    \begin{subfigure}[b]{0.2\columnwidth}
        \includegraphics[width=\columnwidth]{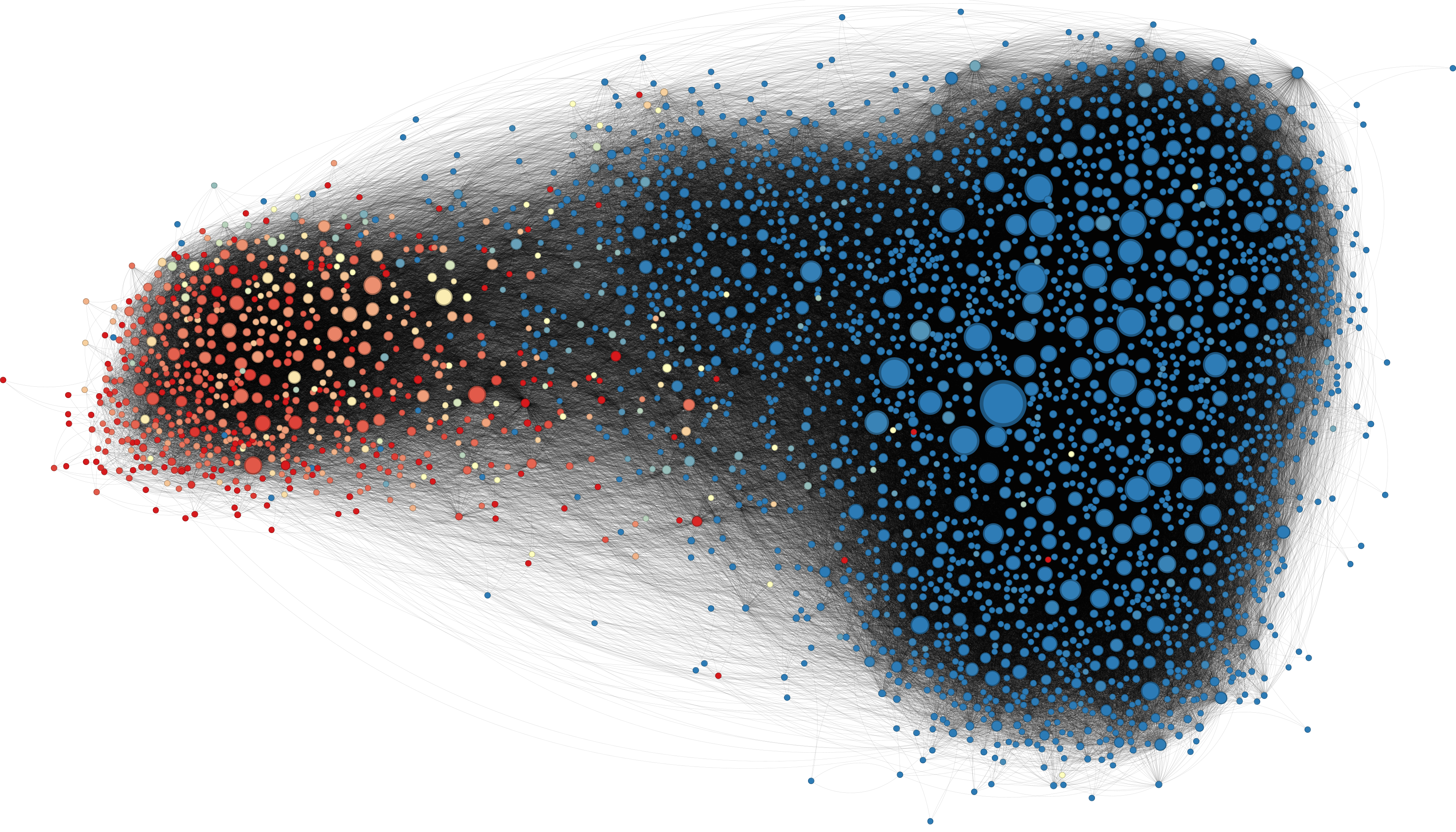} 
        \caption{\scriptsize{Italian Referendum}}
        \label{fig:referendum-viz}
    \end{subfigure}
    \caption{Real-world social networks from $\mathbb{X}$. Nodes colour span from blue ($-1$: ``Remain''/``No'') to red ($1$: ``Leave''/``Yes''), while their size resembles their degree.}
    \label{fig:real_networks}
\end{figure}

\begin{figure}[!ht]
    \centering
    \begin{minipage}{0.4\linewidth}
    \includegraphics[width=\linewidth]{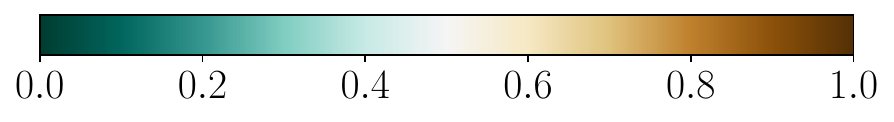}
    \end{minipage}\\
    \rotatebox[origin=c]{90}{\textbf{\textcolor{white}{cia}\footnotesize{Brexit}}}%
    \includegraphics[angle=-25, width=0.2\linewidth]{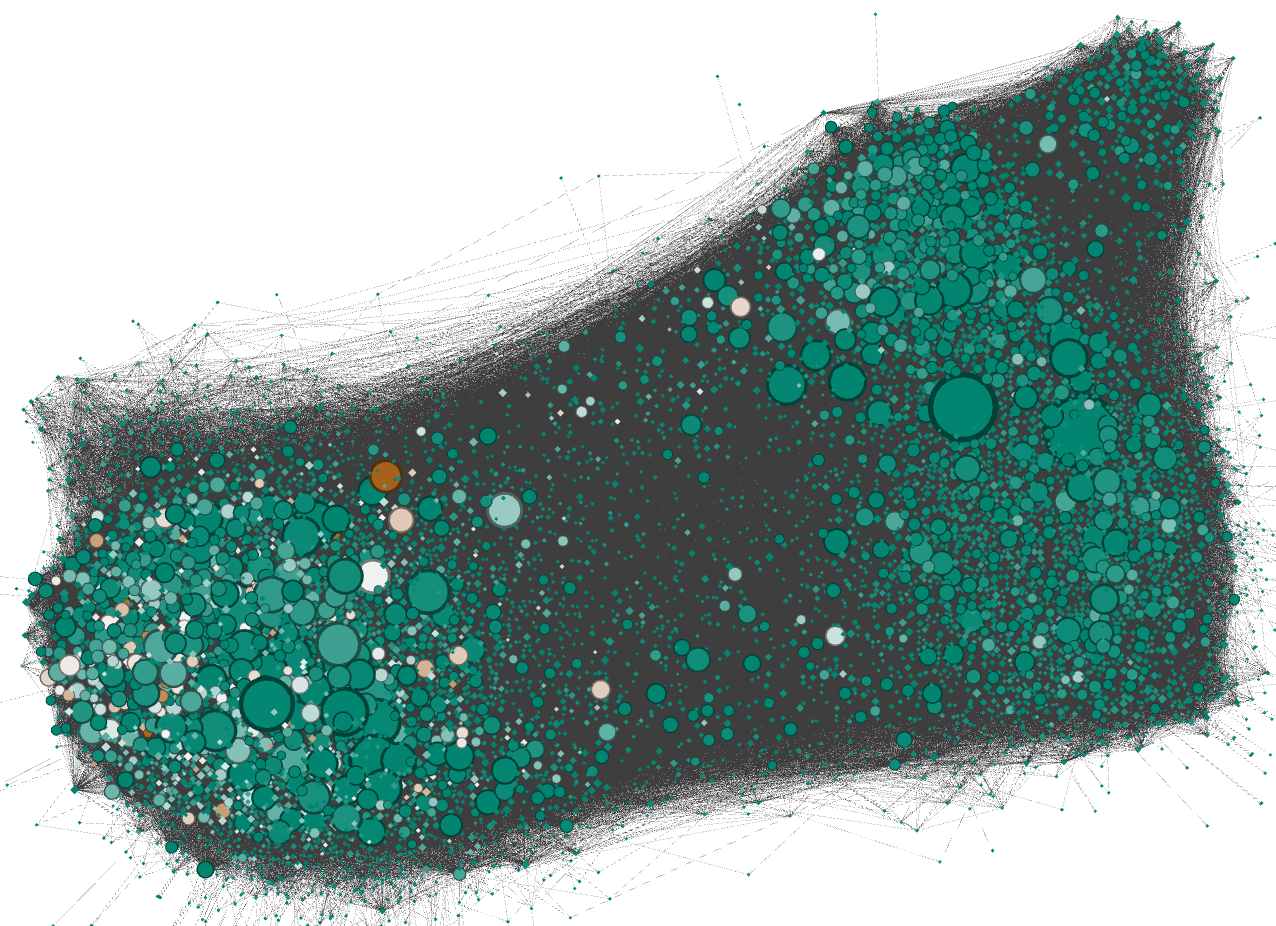}
    \includegraphics[angle=-25, width=0.2\linewidth]{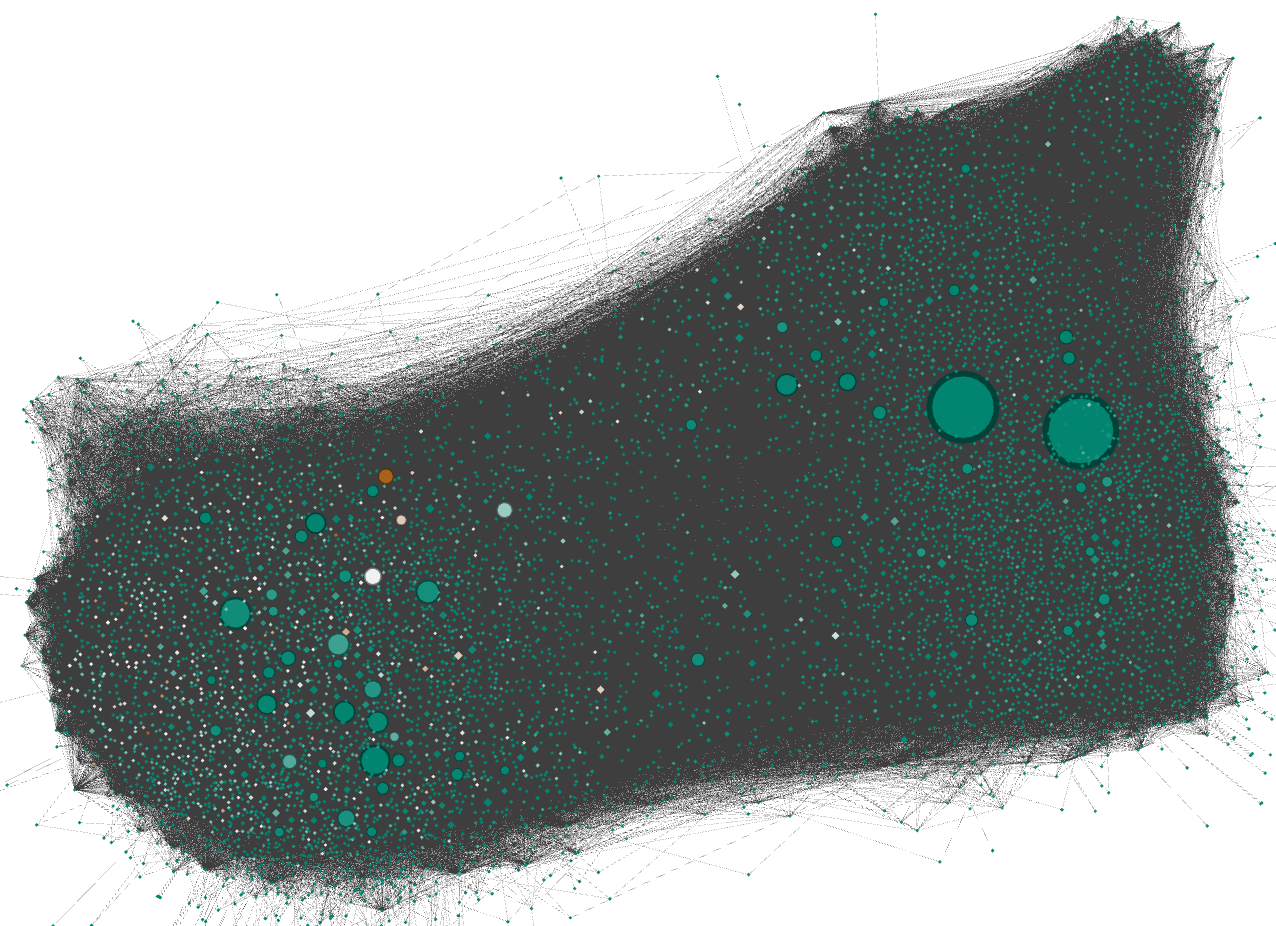}
    \includegraphics[angle=-25, width=0.2\linewidth]{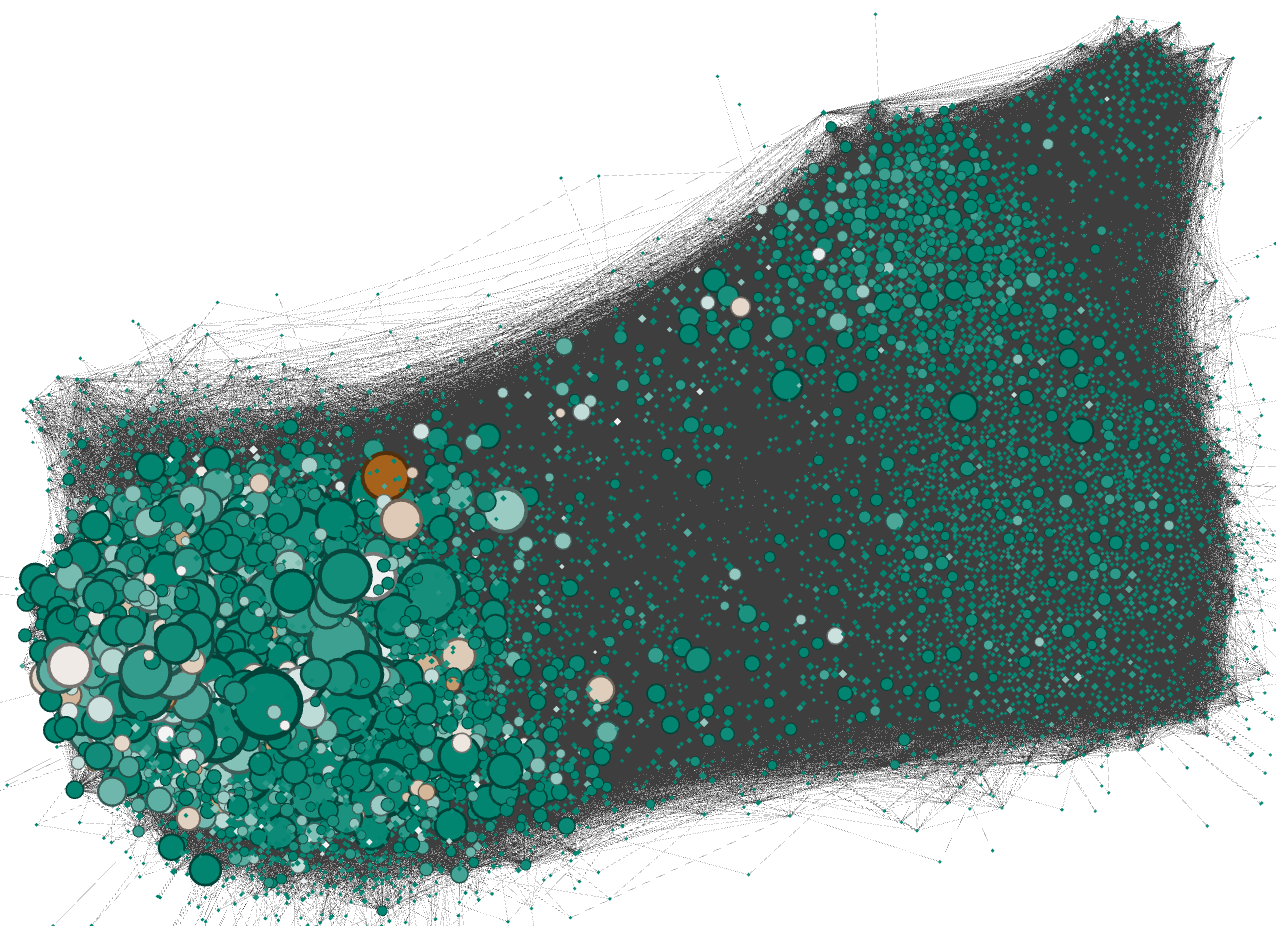}
    \\
    \rotatebox{90}{
    \textbf{\textcolor{white}{ciaooo}\footnotesize{Referendum}}}
    \begin{subfigure}[b]{0.2\linewidth}
    \includegraphics[angle=20, width=\linewidth]{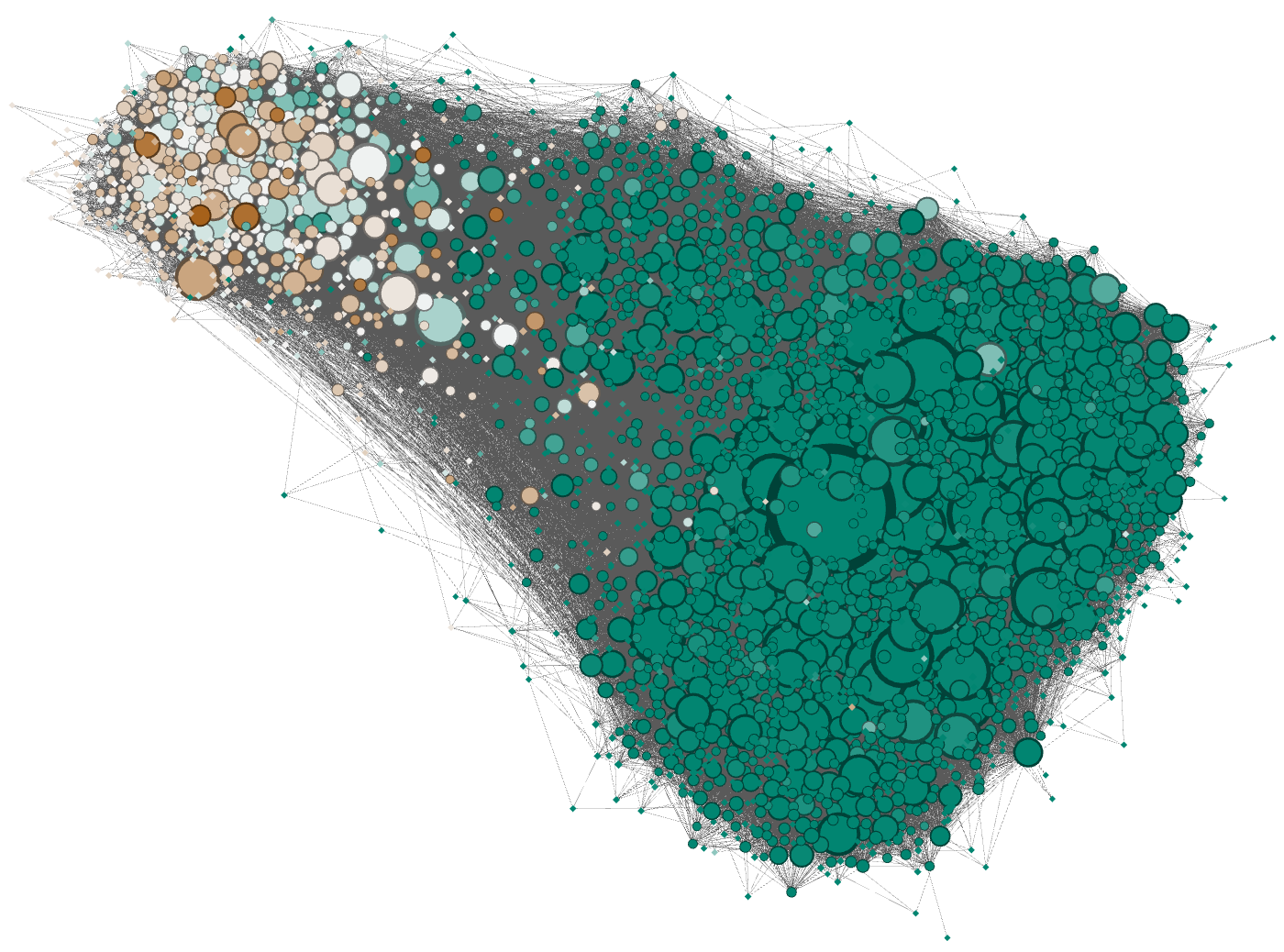}
    \caption{\scriptsize{Degree}\\
    \textcolor{white}{Ciaociao}}
    \end{subfigure}
    \begin{subfigure}[b]{0.2\linewidth}
    \includegraphics[angle=20, width=\linewidth]{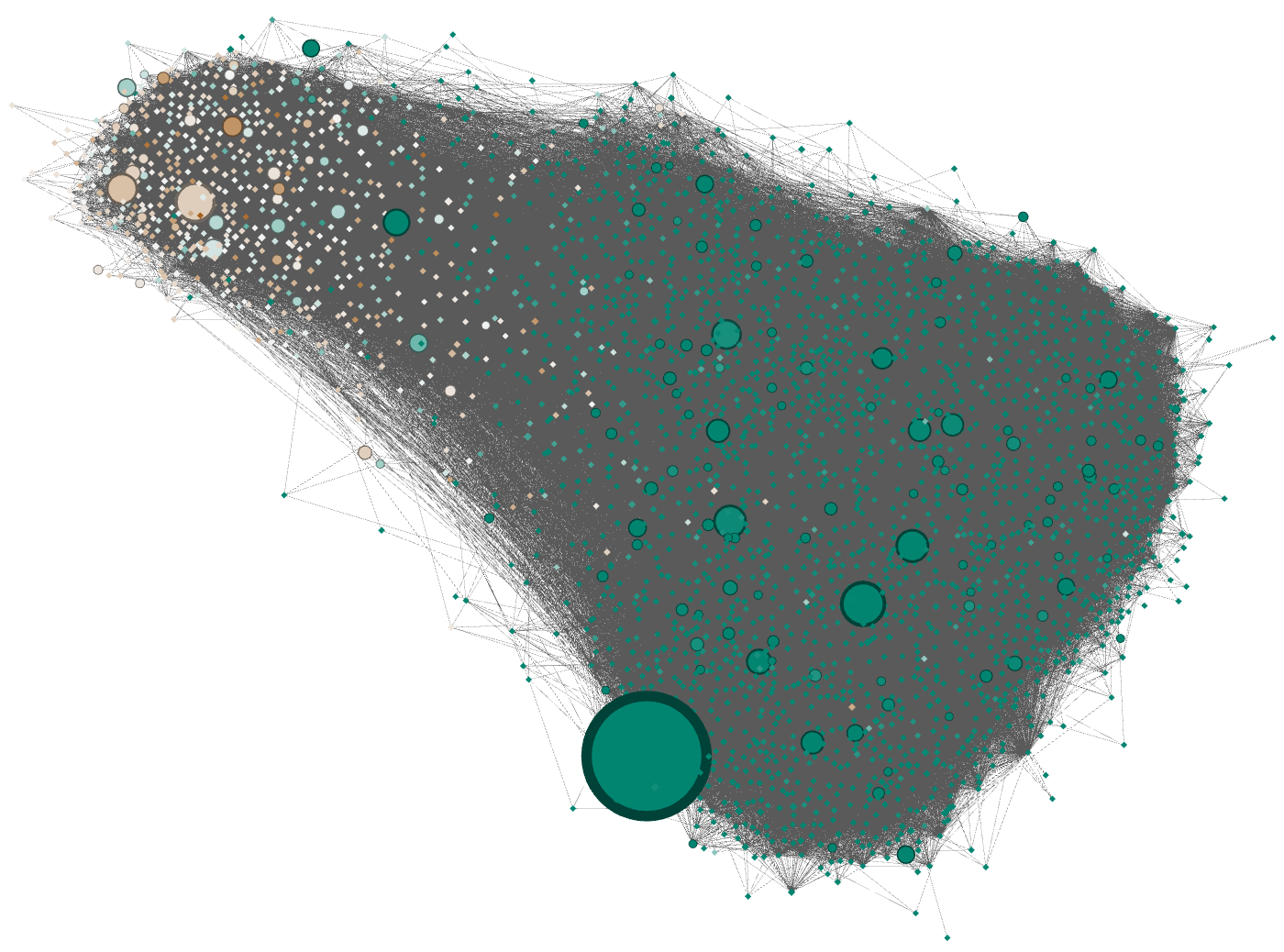}
    \caption{\scriptsize{Betweenness}\\
    \textcolor{white}{ccc}}
    \end{subfigure}
    \begin{subfigure}[b]{0.2\linewidth}
    \includegraphics[angle=20, width=\linewidth]{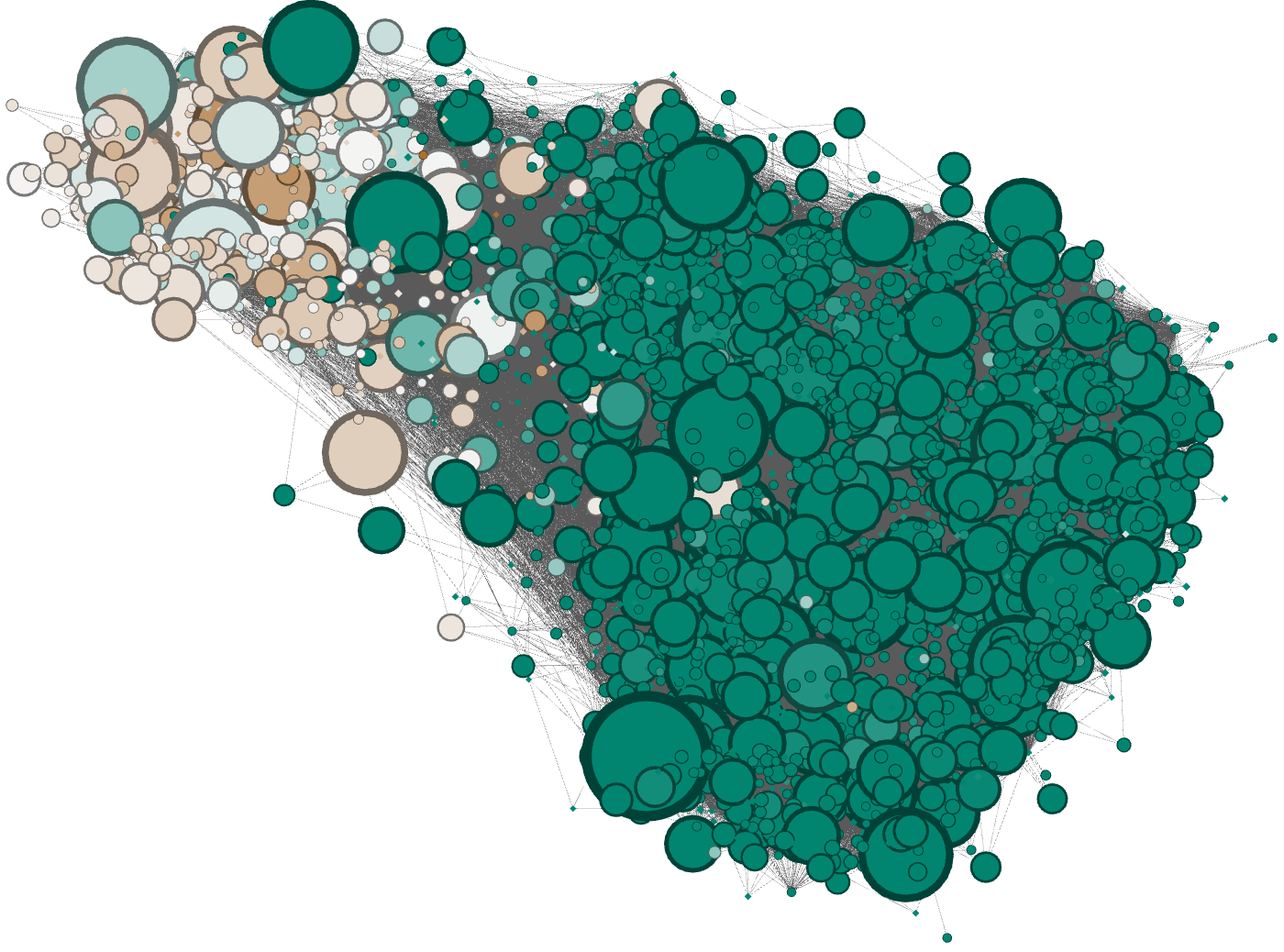}
    \caption{\scriptsize{Eigenvector}\\
    \textcolor{white}{ccc}}
    \end{subfigure}
    \caption{Visualization of the social graphs of Brexit (upper row) and Italian Referendum (bottom row), where nodes are colored according to the induced (normalized) disruption while their size depends on the given centrality measure.}
    \label{fig:disruption-centrality}
\end{figure}

\begin{algorithm}[!ht]
\caption{Fine-tuning framework}
\small
\label{alg:procedure}

\flushleft\textbf{Input:} $\textrm{LLM}_{\theta}$, \innateopinions, $\node$, $\alpha$, $\epsilon$, \prompt, $\mathcal{S}$, $\sigma$, $\kappa$, $\tau$ \\
\textbf{Output:} $\content^*$
\begin{algorithmic}[1]
\State $x =$ None
\State $j = 0, \kappa = 0$
\State $\theta^{(0)} \gets \theta$
\State $\innateopinions' \gets \innateopinions - \alpha \mathbf{e}_\node$ \Comment{Eq.~\ref{eq:s_first}}

\While{$x \not\approx \innateopinion'_\node$ \textbf{and} $\kappa < \tau$}
 \State $\content \gets \textrm{LLM}_{\theta^{(j)}}(\prompt)$
 \State $\kappa\gets \mathbb{E}_{\content\sim P_{\theta^{(j)}}(\cdot|q)}\left[\log\frac{P_{\theta^{(j)}}(\content|q)}{P_{\theta^{(0)}}(\content|q)}\right]$
\State $x \gets \mathcal{S}(\content)$
\State $\mathcal{R} \gets e^{-\frac{(x - s'_u)^2}{2\sigma^2}}$ \Comment{Eq.~\ref{eq:reward}}
\If{$x \approx \innateopinion'_\node$} \Comment{$x \approx \innateopinion'_\node$ if $x \in [\innateopinion'_\node - \epsilon, \innateopinion'_\node + \epsilon]$}
\State $\content^* \gets c$ 
\State \textbf{break}
\EndIf
\State Compute $\theta^{(j+1)}$ by updating $\theta^{(j)}$ and using $\mathcal{R}$ as reward
\State $j \gets j + 1$
\EndWhile
\end{algorithmic}
\end{algorithm}

\begin{table}[!ht]
    \centering
    \caption{Person correlation value $\rho$ between the induced disruption and the nodes centrality measures, computed over Brexit and Italian Referendum real-world graphs.}
    \label{tab:correlation}
    \resizebox{\columnwidth}{!}{
    \begin{tabular}{cccc}
    \toprule
        $\rho$ & Degree & Betweenness Centrality & Eigenvector Centrality \\
        \midrule
        Brexit & 0.135 & 0.03 & 0.275 \\
        Referendum & -0.099 & -0.003 & -0.307 \\
        \bottomrule
    \end{tabular}
    }
\end{table}
Algorithm~\ref{alg:procedure} depicts the pseudo-code of our fine-tuning procedure devised to train the LLM to generate disruptive content.

Figure~\ref{fig:real_networks} depicts the real-world social graphs, while Figure~\ref{fig:disruption-centrality} provides a visualization of the networks where nodes are colored based on their disruption score, while their size depends on the given centrality measure. Disruption has been normalized via min-max scaling. 
The results on the correlation analysis between node centrality and induced disruption are further depicted in Table~\ref{tab:correlation}, where the $\rho$ value corresponds to the Pearson correlation coefficient~\cite{ref1}.

\clearpage

\section*{GenAI Usage Disclosure}
The authors used large language model-based tools for light editing assistance, such as grammar checking and sentence-level rephrasing, in the preparation of this manuscript. All scientific content, experimental design, data collection, analysis, and conclusions are exclusively the work of the human authors. No generative AI tool was used to produce figures, tables, or the benchmark itself.

\bibliographystyle{ACM-Reference-Format}
\bibliography{ref}


\end{document}